\documentclass[12pt]{article}

\usepackage{amsmath,amssymb,epsfig,amscd}

\numberwithin{equation}{section}       % equation numbers in each section
\newcommand{\me}{\mathrm{e}}
\newcommand{\mi}{\mathrm{i}}

\newcommand{\dd}{\mathrm{d}}
\newcommand{\del}{\partial}

\newcommand{\bbZ}{\mathbb{Z}}
\newcommand{\bbR}{\mathbb{R}}
\newcommand{\bbC}{\mathbb{C}}
\newcommand{\bbQ}{\mathbb{Q}}
\newcommand{\bbP}{\mathbb{P}}

\DeclareMathOperator{\SU}{\mathit{SU}}
\DeclareMathOperator{\SO}{\mathit{SO}}
\DeclareMathOperator{\Symp}{\mathit{Sp}}
\DeclareMathOperator{\Spin}{\mathit{Spin}}

\DeclareMathOperator{\SL}{\mathit{SL}}
\DeclareMathOperator{\GL}{\mathit{GL}}

\newcommand{\id}{\mathrm{id}}

\DeclareMathOperator{\vol}{vol}

\DeclareMathOperator{\Cliff}{Cliff}
\DeclareMathOperator{\End}{End}
\DeclareMathOperator{\ad}{ad}
\DeclareMathOperator{\coker}{Coker}
\DeclareMathOperator{\Hol}{Hol}
\DeclareMathOperator{\Ric}{Ric}

\newcommand{\Ricci}[1]{\Ric_{#1}}
\newcommand{\CP}{\mathbb{C}P}
\newcommand{\Csec}{C^\infty}
\newcommand{\AdS}{\mathrm{AdS}}
\newcommand{\rest}[1]{\left.#1\right|}

\newcommand{\rr}{\rho}

\begin{document}

%%%%%%%%%%%%%%%%%%%%%%%%%%%%%%%%%%%%%%%%%%%%%%%%%%%%%%%%%%%%%%%%%%%%%%%%%%

\begin{titlepage}

\vfill

\begin{flushright}
\begin{small}
   Imperial/TP/041101\\ CERN-PH-TH/2004-225\\HUTP-04/A0048\\hep-th/0411194\\
   \end{small}
\end{flushright}

\vfill
\vfill

\begin{center}
   \baselineskip=16pt
   \begin{LARGE}
      \textsl{Supersymmetric AdS Backgrounds\\*[0.6em]
        in String and M-theory}
   \end{LARGE}
   \vskip 2cm
      Jerome P. Gauntlett$^{1}$, Dario Martelli$^{2}$, James
      Sparks$^{3}$ and Daniel Waldram$^{1}$
   \vskip .6cm
   \begin{small}
      \textit{$^{1}$Blackett Laboratory, Imperial College\\
        London, SW7 2AZ, U.K.}
        %E-mail: j.gauntlett, d.waldram@imperial.ac.uk}
        \end{small}
 \vskip .6cm
   \begin{small}
      \textit{$^{2}$Department of Physics, CERN Theory Division\\
        1211 Geneva 23, Switzerland}
        %E-mail: dario.martelli@cern.ch}
        \end{small}
\vskip .6cm
   \begin{small}
      \textit{$^{3}$Department of Mathematics, Harvard University\\
        One Oxford Street, Cambridge, MA 02318, U.S.A.\\
    {\it and}\\
    Jefferson Physical Laboratory, Harvard University\\
    Cambridge, MA 02318, U.S.A.}
        %E-mail: sparks@math.harvard.edu}
   \end{small}
\end{center}

\vfill

\begin{small}

\begin{center}
   \textbf{Abstract}
\end{center}

\begin{quote}
We first present a short review of general supersymmetric
compactifications in string and M-theory using the language of
$G$-structures and intrinsic torsion.
We then summarize recent work on the generic conditions for
supersymmetric $\AdS_5$ backgrounds in M-theory and the
construction of classes of new solutions. Turning to $\AdS_5$
compactifications in type IIB, we summarize the construction of an
infinite class of new Sasaki--Einstein manifolds in dimension
$2k+3$ given a positive curvature K\"ahler--Einstein base manifold
in dimension $2k$. For $k=1$ these describe new supergravity duals
for ${\cal N}=1$ superconformal field theories with both rational
and irrational R-charges and central charge. We also present a
generalization of this construction, that has not appeared
elsewhere in the literature, to the case where the base is a
product of K\"ahler--Einstein manifolds.

\vfill
\vfill

\textit{Based on a talk by DW at the 73rd Meeting between Theoretical
  Physicists and Mathematicians, ``The (A)dS-CFT correspondence'',
  Strasbourg, France.}
\end{quote}
\end{small}

\end{titlepage}

%%%%%%%%%%%%%%%%%%%%%%%%%%%%%%%%%%%%%%%%%%%%%%%%%%%%%%%%%%%%%%%%%%%%%%%%%%

\section{Introduction}
\label{sec:intro}

In this paper we aim to review first the general framework of
supersymmetric solutions of string or M-theory, where spacetime is
a product $E\times X$ of an external manifold $E$ and an internal
manifold $X$, and then, secondly, two interesting classes of
examples where $E$ is five-dimensional anti-de Sitter ($\AdS_5$)
space and $X$ is five- or six-dimensional. This latter work was
first presented in three papers, refs.~\cite{m6},~\cite{se}
and~\cite{gse}. Such backgrounds are central in string theory
first, when $E$ is four-dimensional Minkowski space, as a way to
construct semi-realistic supersymmetric models of particle
physics, and second, when $E$ is an AdS space, as gravitation
duals of quantum conformal field theories, via the AdS-CFT
correspondence (for a review see ref.~\cite{adscft-rev}).

We consider string or M-theory in the low-energy supergravity
limit where the condition for a supersymmetric solution requires
the existence of a constant spinor with respect to a particular
Clifford algebra-valued connection $D^X$, perhaps supplemented
with additional algebraic conditions on the spinor. When certain
fields, so-called $p$-form fluxes, in the supergravity are zero,
$D^X$ is equal to the Levi--Civita connection and hence
supersymmetry translates into a condition of special holonomy.
However, in many cases one wants to include non-trivial flux. In
the first part of the paper we review how this translates into the
existence of a $G$-structure $P$ and how the fluxes are encoded in
the intrinsic torsion of $P$. We also comment on the relation to
generalized holonomy and generalized calibrations. By way of an
example we concentrate on the case of $d=11$ supergravity on a
seven-dimensional $X$ with $\SU(3)$-structure, and type IIB
supergravity on a six-dimensional $X$ and only five-form flux with
an $\SU(3)$-structure.

The second part of the paper, based on ref.~\cite{m6}, discusses first
the general conditions on the geometry of $X$ in supersymmetric
$\AdS_5\times X$ solutions of $d=11$ supergravity, and second a
large family of explicit regular solutions of this form
characterized by $X$ being complex. Previously, a surprisingly
small number of explicit solutions were known. Most notable was
that of Maldacena and Nu\~nez~\cite{malnun} describing the near
horizon limit of fivebranes wrapping constant curvature
holomorphic curves in Calabi--Yau three-folds. The new solutions
can be viewed as corresponding to a more general type of embedded
holomorphic curve. They fall into two classes where $X$ is a
fibration of a two-sphere over either a four-dimensional
K\"ahler--Einstein manifold or a product of constant-curvature
Riemann surfaces.

The third part of the paper relates to Sasaki--Einstein (SE)
manifolds. These arise as the internal manifold $X$ in
supersymmetric type IIB $\AdS_5\times X$ solutions. We review a
new construction~\cite{se,gse} of an infinite class of SE
manifolds in any dimension $n=2k+3$ based on an underlying
$2k$-dimensional positive curvature
K\"ahler--Einstein manifold. All SE spaces have a
constant norm Killing vector $K$ (see, for instance,
refs.~\cite{sas-rev} and~\cite{dnp}) and can be characterized by
whether the orbits of $K$ are closed (so-called regular and
quasi-regular cases) or not (the irregular case). The new class of
solutions includes quasi-regular and irregular cases. Again,
previously, surprisingly few explicit SE metrics were known: the
homogeneous regular cases have been classified~\cite{tri}; several
quasi-regular examples had been constructed using algebraic
geometry techniques but without an explicit metric; and no
irregular examples were known. Finally we give a straightforward
extension of the construction to the case where the underlying
manifold is a product of K\"ahler--Einstein spaces. This is new
material and leads to new $AdS_4\times X_7$ solutions of M-theory.

\hfill \hfill * \hfill * \hfill * \hfill \hfill \

The history of considering supersymmetric backgrounds of
supergravity theories with non-trivial fluxes is a comparatively
long one. The use of $G$-structures to classify such backgrounds
was first proposed in ref.~\cite{Gauntlett:2002sc}. This was based
partly on earlier work by Friedrich and Ivanov~\cite{fr-iv},
though these authors did not consider the supergravity equations
of motion. The relationship between background supersymmetry
conditions and generalized calibrations~\cite{g-cal} was first
discussed slightly earlier in ref.~\cite{Gauntlett:2001ur} and shown
to be generic in ref.~\cite{11d,int-tor}.

These techniques have subsequently been developed and extended in
a number of directions. First, one can use $G$-structures to classify
{\it all} supersymmetric solutions of a given supergravity
theory. This has been carried out for the most 
generic case of a single preserved supersymmetry in $d=11$
supergravity in ref.~\cite{11d}. A similar classification has now
also been carried out for simpler supergravity theories in
four~\cite{4d}, five~\cite{5d}, six~\cite{6d} and seven~\cite{7d} 
dimensions. Note that this work extends older work of Tod~\cite{Tod}
which classified supersymmetric solutions of four-dimensional
supergravity using techniques specific to four dimensions.
An important open problem in, for instance, the $d=11$ case is to
refine the classification presented in ref.~\cite{11d} and determine
the extra conditions required for solutions to preserve more than one
supersymmetry. There has been some recent progress on this using
$G$-structures, partly implementing some suggestions in 
ref.~\cite{11d}, in the context of seven~\cite{oisin} and
eleven dimensions~\cite{ggp}. Note that the case of maximal
supersymmetry can be analysed using different techniques and this
has been carried out for type IIB and $d=11$ supergrvaity in
ref.~\cite{Figueroa-O'Farrill:2002ft}. A quite different attempt at
classification, first advocated in refs.~\cite{genhol} and
subsequently studied in ref.~\cite{genholall}, is to use the notion of
``generalized holonomy''. We comment on the relation to $G$-structures
in the next section. 

A second application is to use $G$-structures to analyse
supersymmetric ``flux compactifications'' in string theory. These
are supersymmetric backgrounds where the external space $E$ is
flat Minkowski space and $X$ is often, but not always, compact.
This is a large field with a rich literature, starting with that
of Strominger~\cite{strom} and Hull~\cite{hull} in the context of
the heterotic string (see also ref.~\cite{wsh}). More recently,
starting with Polchinski and Strominger~\cite{ps} as well
as ref.~\cite{bachas}, several authors have analyzed flux backgrounds
for the special case where $X$ is a special holonomy manifold (for 
early work see refs.~\cite{bb}--\cite{tv}) and the resulting
low-energy effective theories on $E$ (a large field, see, for
instance, the references in ref.~\cite{Gurrieri:2002wz} or
ref.~\cite{Cardoso:2002hd}). Let us concentrate on the use of
$G$-structure techniques to analyse cases when $X$ does not have
special holonomy . In ref.~\cite{Gurrieri:2002wz} it was argued
that the mirror of a Calabi--Yau threefold with three-form
$H$-flux is a manifold with a ``half-flat'' $\SU(3)$-structure.
Further work in this direction appears in refs.~\cite{minas}. For the
heterotic string, Strominger's and Hull's results imply $X$ has a
non-K\"ahler $\SU(3)$ structure and these have been analyzed
in refs.~\cite{int-tor}, \cite{Cardoso:2002hd}--\cite{Ivanov:2003nd}.
Such flux compactifications have only $H$-flux, and these were
completely classified, including type II backgrounds, in
ref.~\cite{int-tor}. (Note that ref.~\cite{int-tor} corrects a sign
in ref.~\cite{strom}, disqualifying the putative Iwasawa solutions
in ref.~\cite{Cardoso:2002hd}.) Flux compactifications on more general
$\SU(3)$-structures have been considered in
refs.~\cite{Gurrieri:2002iw}--\cite{Micu:2004tz}. General type II
compactifications with more general fluxes have been addressed, for
instance, in refs.~\cite{Grana:2001xn}--\cite{Behrndt:2004mj}. General
$d=11$ flux compactifications have been discussed in terms of 
$G$-structures in several papers~\cite{Dall'Agata:2003ir},
\cite{Kaste:2003dh}--\cite{House:2004hv}.

A third connected application is to spacetime solutions dual to
supersymmetric field theories via the AdS-CFT correspondence. The
basic case of interest is when $E$ is AdS since the solution is
then dual to a supersymmetric conformal field theory. There are
more general kinds of solutions, however, that are dual to other
types of field theories, as well as to renormalisation group flows
(see the review~\cite{adscft-rev}). Again this is very large
field. Aside from the initial paper~\cite{Gauntlett:2002sc} (which
focussed on solutions dual to ``little string theories") and the
work~\cite{m6,se,gse} on which this paper is based, $G$-structures
have been used to analyze $\AdS$ solutions in
refs.~\cite{Martelli:2003ki} and~\cite{Lukas:2004ip}. Very
recently an interesting class of half-supersymmetric solutions has
been found~\cite{llm}. Note that there is also a related approach
to finding special sub-classes of solutions initiated by Warner
and collaborators (see for instance refs.~\cite{warner}).

Finally, we comment on some work related to $G$-structure
classifications and generalised calibrations. Calibrations are
important in string theory backgrounds with vanishing fluxes since
the calibrated cycles are the cycles static probe branes can wrap
whilst preserving supersymmetry. Generalised calibrations~\cite{g-cal}
are the natural generalisation to backgrounds when the fluxes are
non-vanishing. Important work relating calibrations and the
superpotential of the effective theory on $E$ first appeared in
ref.~\cite{calW}. Starting with the work~\cite{Gauntlett:2001ur} and
subsequent work including refs.~\cite{11d,int-tor,Martelli:2003ki} it
has become clear that the conditions placed on supersymmetric
backgrounds often have the useful physical interpretation as generalised
calibrations. The reason for this is simply that the backgrounds
can arise when branes wrap calibrated cycles after taking into
account the back-reaction (see ref.~\cite{int-tor} for further
discussion). Such wrapped brane solutions were first found in
ref.~\cite{malnun} and a review can be found in
ref.~\cite{jerome}. The relationship between wrapped and intersecting
brane solutions and generalised calibrations has been studied in 
refs.~\cite{kastor,Gauntlett:2001qs,husain}. The classifcation of
supersymmetric solutions using $G$-structures has also led to a
further exploration of generalised calibrations for non-static
brane configurations~\cite{smith}. Further work, specifically on the
relationship between supersymmetry and generalized calibrations in
flux compactifications, has appeared in
refs.~\cite{Gauntlett:2001ur,int-tor,angel}.

%%%%%%%%%%%%%%%%%%%%%%%%%%%%%%%%%%%%%%%%%%%%%%%%%%%%%%%%%%%%%%%%%%%%%%%%%%

\section{Supersymmetry and $G$-structures}
\label{sec:gen}

\subsection{Some supergravity}
\label{sec:sugra}

Let us start by characterising the type of problem we are trying to
solve. First we summarise a few relevant parts of the supergravity
theories which arise in string theory and then describe the notion of
a supersymmetric compactification or reduction. We will concentrate on
two fairly generic examples in ten and eleven dimensions.

We start with a \emph{supergravity theory} on a $d$-dimensional
Lorentzian spin manifold $M$. This is an approximation to the full string
theory valid in the limit where the curvature of the manifold is small
compared with the intrinsic string scale. The supergravity is
described in terms a number of fields, including the \emph{bosonic}
fields
\begin{equation}
\label{eq:fields}
\begin{aligned}
   g &   && \text{Lorentzian metric} , \\
   \Phi &\in \Csec(M) && \text{dilaton} ,\\
   F^{(p)} &\in \Csec(\Lambda^pT^*M) && \text{$p$-form fluxes} ,
\end{aligned}
\end{equation}
for certain values of $p$, satisfying equations of motion which
are generalisations of Einstein's and Maxwell's equations.
Particular $p$-form fluxes are also sometimes labeled $G$ or $H$.
For the cases we will consider, the dilaton $\Phi$ is either not
present in the theory or assumed to be zero. Since the theory is
supersymmetric these fields are paired with a set of
\emph{fermionic} fields transforming in spinor representations.
However, these will all be set to zero in the backgrounds we
consider. A bosonic solution of the equations of motion is called
a supergravity \emph{background}.

We would like to characterise \emph{supersymmetric backgrounds}.
Let $S\to M$ be a spin bundle. (Precisely which spinor
representation we have depends on the dimension $d$ and the type
of supergravity theory.) The supergravity theory defines a
particular connection
\begin{equation}
   D : \Csec(S)\to \Csec(S\otimes T^*M)
      \qquad \text{supergravity connection} ,
\end{equation}
in terms of the metric, dilaton and $p$-form fluxes. A background is
supersymmetric if we have a non-trivial solution to
\begin{equation}
\label{genKS}
   D \epsilon = 0
      \qquad \text{Killing spinor equation} ,
\end{equation}
for $\epsilon\in\Csec(S)$. If we have $n$ independent solutions then
the background is said to preserve $n$ supersymmetries. (Often the
supergravity also defines a map $P\in\Csec(\End(S))$ in terms of the
dilaton and $F^{(p)}$ and a supersymmetric solution must
simultaneously satisfy the ``dilatino equation'' $P\epsilon=0$. For
our particular examples either $P$ is not present in the
supergravity theory or is assumed to be identically zero.)

The two cases we will consider are (1)~$d=11$ supergravity with
four-form flux $G$ and (2)~$d=10$ \emph{Type IIB} supergravity keeping
only a self-dual five-form flux $F^{(5)}=*F^{(5)}$. The
corresponding supergravity connections are given, in components, by
\begin{align}
   D &= \nabla^g + \tfrac{1}{12}G\lrcorner\Gamma^{(5)}
      + \tfrac{1}{6}\Gamma^{(3)}\lrcorner G
      && \text{$d=11$} , \label{DM} \\
   D &= \nabla^g \otimes \id
      - \tfrac{1}{8} \Gamma^{(4)}\lrcorner F^{(5)}\otimes \mi\sigma_2
      && \text{Type IIB} , \label{D2B}
\end{align}
where $\nabla^g$ is Levi--Civita connection for $g$. In the first
case, the spinor $\epsilon$ is a 32-dimensional real
representation $\Delta^\bbR_{10,1}$ of $\Spin(10,1)$ while in the
second case $\epsilon$ is a \emph{pair} of spinors
$(\epsilon_1,\epsilon_2)$ each in the 16-dimensional real, chiral
representation $\Delta^{+\bbR}_{9,1}$ of $\Spin(9,1)$. The gamma
matrices $\Gamma$ generate $\Cliff(d-1,1)$ and $\Gamma^{(p)}$ is
the antisymmetrised product of $p$ gamma matrices. The matrices
$\id=\left(\begin{smallmatrix}1&0\\0&1\end{smallmatrix}\right)$
and
$\mi\sigma_2=\left(\begin{smallmatrix}0&1\\-1&0\end{smallmatrix}\right)$
act on the doublet of spinors
$\epsilon=\left(\begin{smallmatrix}\epsilon_1\\ \epsilon_2
\end{smallmatrix}\right)$.  In index notation we have $(v\lrcorner
w)_{M_{p+1}\dots M_{q}}=\frac{1}{p!}v^{M_1\dots M_p}w_{M_1\dots
M_pM_{p+1}\dots M_{q}}$.

%%%%%%%%%%%%%%%%%%%%%%%%%%%%%%%%%%%%%%%%%%%%%%%%%%%%%%%%%%%%%%%%%%%%%%%%%%

\subsection{The problem}
\label{sec:problem}

It is interesting to determine what the existence of solutions to
the Killing spinor equation implies about the geometry of $M$, in
general. For example, this has been studied in ref.~\cite{11d} for the
most general solutions of supergrvaity in eleven dimensions.
However, here we are concerned with a more restricted problem.

First we assume we have a \emph{compactification}, where the
topology of $M$ is taken to be a product
\begin{equation}
   M = E \times X
\end{equation}
of a $(d-n)$-dimensional \emph{external} manifold $E$ and an
$n$-dimensional \emph{internal} manifold  $X$. Although compact
$X$ is often of most interest, by an abuse of terminology, we will
also allow for non-compact $X$. Next, the metric is taken to be a
warped product. In particular we consider two cases
\begin{equation}
\label{eq:geom}
\begin{aligned}
   g &= \me^{2\Delta} \eta_{d-n} + g_{X} && \text{flat space}, \\
   g &= \me^{2\lambda} \phi_{d-n} + g_{X} && \text{AdS space}, \\
\end{aligned}
\end{equation}
where $g_X$ is a Riemannian metric on $X$ while $\eta_r$ is the flat
Minkowski metric on $E=\bbR^{r-1,1}$ and $\phi_r$ is the constant
curvature metric on anti-de~Sitter space $E=\AdS_r$. In the latter
case $\Ricci{\phi_r}=-(r-1)m^2\phi_r$ where $m$ is the inverse radius
of the AdS space. In each case
\begin{equation}
   \lambda, \Delta \in \Csec(X) .
\end{equation}
Finally the dilaton and fluxes are assumed to be given by objects on
$X$, so that in the two cases we have
\begin{equation}
\label{flux-ansatz}
\begin{aligned}
   F^{(p)} &= \begin{cases}
         \me^{(d-n)\Delta}\vol_{\eta_{d-n}}\wedge\, f + h , \\
         \me^{(d-n)\lambda}\vol_{\phi_{d-n}}\wedge\, f + h ,
         \end{cases}
      \qquad
      f \in \Csec(\Lambda^{n-d+p}T^*X),\
      h \in \Csec(\Lambda^pT^*X), \\
   \Phi &\in \Csec(X).
\end{aligned}
\end{equation}
where $\vol_{g_E}$ is the volume form corresponding to the metric
$g_E\in\{\eta_r,\phi_r\}$ on $E$ and $f$ and $h$ are sometimes
referred to the as the electric and magnetic fluxes.

Physically such solutions are interesting because, first, in the
flat-space case with $d-n=4$, the space $E$ is a model for
four-dimensional particle physics. Secondly, the AdS-CFT
correspondence~\cite{adscft-rev} implies that such AdS geometries
should be gravitational duals of conformal field theories in
$d-n-1$ dimensions. The particular internal $X$ geometry encodes
the content of the particular conformal field theory.

Given this product ansatz the Killing spinor
equation~\eqref{genKS} reduces to equations on a spinor $\psi$ of
$\Spin(d-n-1,1)$ on $E$ and a spinor, not necessarily irreducible,
$\xi$ of $\Spin(n)$ on $X$.  The exact decomposition of
$\epsilon\in\Csec(S)$ depends on the dimensions $d$ and $n$. In
all cases one takes $\psi$ to satisfy the standard Killing spinor
equation on $E$, that is
\begin{equation}
\label{E-KS}
\begin{aligned}
   \nabla^{\eta}\psi &= 0 && \text{flat space} , \\
   \left(\nabla^{\phi}-\tfrac{1}{2}m\rho\right)\psi &= 0
      && \text{AdS space} ,
\end{aligned}
\end{equation}
where $\nabla^\eta$ and $\nabla^\phi$ are the Levi--Civita connections
for $\eta_{d-n}$ and $\phi_{d-n}$ respectively and $\rho$ are gamma
matrices for $\Spin(d-n-1,1)$. If $S^X\to X$ is the spin bundle on $X$
coming from the decomposition of $S$, the Killing spinor
equation~\eqref{genKS} then has the form\footnote{If there was also
  originally a $P\epsilon=0$ condition this also reduces to a further
  condition $P^X\xi=0$ with $P^X\in\Csec(\End(S^X))$.}
\begin{equation}
\label{KS}
   D^X\xi = 0,  \qquad Q^X\xi = 0,
      \qquad \text{reduced Killing spinor eqns.} ,
\end{equation}
where the connection $D^X:\Csec(S^X)\to\Csec(S^X\otimes T^*X)$ and the
map $Q^X:\Csec(S^X)\to\Csec(S^X)$ each are defined in terms of flux,
dilaton, and $\Delta$ or $\lambda$ and $m$. The condition $D^X\xi=0$
comes from the reduction of $D\epsilon=0$ on $X$ and $Q^X\xi=0$ from
the reduction on $E$.

Our basic question is then
\begin{quote}
   \textit{what does the existence of solutions to the reduced Killing spinor
   equations imply about the geometry of $X$ and the form of the
   fluxes and dilaton?}
\end{quote}
In general we want to translate the Killing spinor conditions into
some convenient set of necessary and sufficient conditions, such as,
for instance, $X$ has a particular almost complex or contact structure
or a particular Killing vector.

Let us end with a couple of further comments. First note that there is
a connection between the two types of
compactification~\eqref{eq:geom}. Consider $E\times
X=\AdS_{d-n}\times X$. Locally we can write the AdS metric
$\phi_{d-n}$ in Poincar\'e coordinates
\begin{equation}
\label{eq:poincare}
   \phi_{d-n} = \me^{-2mr}\eta_{d-n-1} + \dd r\otimes \dd r .
\end{equation}
Thus we have
\begin{equation}
\begin{aligned}
   g = \me^{2\lambda}\phi_{d-n} + g_X
      &= \me^{2\lambda}\me^{-2mr}\eta_{d-n-1}
         + \left(\me^{2\lambda}\dd r\otimes \dd r + g_{X}\right) \\
      &\equiv \me^{2\Delta}\eta_{d-n} + g_{X'} ,
\end{aligned}
\end{equation}
where
\begin{equation}
\label{reduction}
\begin{aligned}
   \Delta &=\lambda-mr , \\
   g_{X'} &=\me^{2\lambda}\dd r\otimes \dd r+g_X .
\end{aligned}
\end{equation}
and hence an AdS compactification on $X$ to $\AdS_{d-n}$ is really
a special case of a flat space compactification on
$X'=X\times\bbR^+$ to $\bbR^{d-n-2,1}$. This will be particularly
useful for deriving the conditions on the geometry of AdS
compactifications in what follows.

Next, recall that to be a true background the fields also have to
satisfy the supergravity equations of motion. Part of these are a
set of Bianchi identities involving the exterior derivatives of
$F^{(p)}$. In general one can derive equations involving the Ricci
tensor and derivatives of the fluxes by considering integrability
conditions, such as $D^2\epsilon=0$, for the Killing spinor
equations. One can show, following ref.~\cite{wsh,11d} that, for
product backgrounds of the form~\eqref{eq:geom}, once one imposes
the Bianchi identities and the equation of motion for the flux the
other equations of  motion follow from these integrability
conditions. In fact, for the cases we consider, the flux equation
of motion is also implied by the supersymmetry conditions and so
if we have a solution of the Killing spinor equation~\eqref{genKS}
and in addition the Bianchi identity
\begin{equation}
\label{BI}
   \dd G = 0 \quad \text{or} \quad \dd F^{(5)}=0
     \qquad \text{Bianchi identity} ,
\end{equation}
then we have a solution of the equations of motion. When $E=\AdS$,
at least for the cases considered here, the supersymmetry
conditions are even stronger: any solution of the Killing spinor
equations is necessarily a solution of the equations of
motion~\cite{m6}.

To have truly a string or M-theory background as opposed to a
supergravity solution there is also a ``quantisation'' condition
on the fluxes. For $n<8$ the equations of motion for $G$ gives
$\dd*G=0$ while $\dd*F^{(5)}=0$ is implied by the Bianchi identity
since $F^{(5)}$ is self-dual. Hence in both cases the fluxes are
harmonic. To be a true string or M-theory background, we have the
quantisation condition $G\in H^4(X,\bbZ)$ or $F^{(5)}\in
H^5(X,\bbZ)$. More precisely the fluxes represent classes in
K-theory~\cite{Ktheory}. In the AdS-CFT correspondence, these
integer classes are related to integral parameters in the field
theory.

%%%%%%%%%%%%%%%%%%%%%%%%%%%%%%%%%%%%%%%%%%%%%%%%%%%%%%%%%%%%%%%%%%%%%%%%%%

\subsection{$G$-structures}
\label{sec:Gstruc}

Our approach for analysing what solutions to the reduced Killing
spinor equations~\eqref{KS} imply about the geometry of $X$ will use
the language of $G$-structures. Let us start with a brief review. For
more information see for instance refs.~\cite{joyce} or~\cite{salamon}.

Let $F$ be the frame bundle of $X$, then
\begin{center}
   a \emph{$G$-structure} is a principle sub-bundle $P$ of $F$ with fibre
   $G\subset\GL(n,\bbR)$.
\end{center}
For example if $G=O(n)$, the sub-bundle is interpreted as the set
of orthonormal frames and defines a metric. Let $\nabla$ be a
connection on $F$ or equivalently the corresponding connection on
$TM$. One finds
\begin{quote}
\begin{enumerate}
\item given a $G$-structure, all tensors on $X$ can be decomposed into
   $G$ representations;
\item if $\nabla$ is \emph{compatible} with the $G$-structure,
that is,
   it reduces to a connection on $P$, then $\Hol(TX,\nabla)\subseteq
   G$;
\item there is an \emph{obstruction} to finding torsion-free
   compatible $\nabla$, measured by the \emph{intrinsic torsion}
   $T_0(P)$, which can be used to classify $G$-structures.
\end{enumerate}
\end{quote}
The intrinsic torsion is defined as follows. Given a pair
$(\nabla',\nabla)$ of compatible connections, viewed as connections
on $P$ we have $\nabla'-\nabla\in\Csec(\ad P\otimes T^*X)$. Let
$T(\nabla)\in\Csec(TX\otimes\Lambda^2T^*X)$ be the torsion of
$\nabla$. We can then define a map $\sigma_P:\Csec(\ad P\otimes
T^*X)\to\Csec(TX\otimes\Lambda^2T^*X)$ given by
\begin{equation}
   \alpha=\nabla'-\nabla \mapsto \sigma_P(\alpha)=T(\nabla')-T(\nabla) ,
\end{equation}
and hence we have the quotient bundle
$\coker\sigma_P=TX\otimes\Lambda^2T^*X/\alpha(\ad P\otimes T^*X)$.
Let the intrinsic torsion $T_0(P)$ be the image of $T(\nabla)$ in
$\coker\sigma_P$ for any compatible connection $\nabla$. By
definition it is the part of the torsion independent of the choice
of compatible connection and only depends on the $G$-structure
$P$.

We will be interested in the particular class of $G$-structures where
\begin{quote}
\begin{enumerate}
\item $P$ can be defined in terms of a finite set $\eta$ of
   $G$-invariant tensors on $X$,
\item $G\subset O(n)$.
\end{enumerate}
\end{quote}
Prime examples of the former condition are an almost complex
structure with $G=\GL(k,\bbC)\subset\GL(2k,\bbR)$, or an
$O(n)$-structure defined by a metric $g$. The sub-bundle of frames
$P$ is defined by requiring the tensors to have a particular form.
For instance, for the $O(n)$-structure we define $P$ as the set of
frames such that the metric $g$ has the form
\begin{equation}
   g = e^1\otimes e^1 + \dots + e^n \otimes e^n .
\end{equation}

These restrictions imply a number of useful results. From the first
condition it follows that
\begin{equation}
   \text{$\nabla$ is compatible with $P$} \quad \Leftrightarrow \quad
      \nabla\Xi=0 , \quad \forall \Xi\in\eta .
\end{equation}
The second condition implies that $P$ defines a metric $g$ and hence
an $O(n)$ structure $Q$. A key point, given $\ad Q\cong\Lambda^2T^*X$,
is that $\sigma_Q$ is in fact an isomorphism and hence
\begin{quote}
   an $O(n)$-structure with metric $g$ has a unique compatible
   torsion-free connection, namely the Levi--Civita connection
   $\nabla^g$.
\end{quote}
Any $P$-compatible connection $\nabla$ can then be written as
$\nabla=\nabla^g+\alpha+\alpha^\perp$ where $\alpha$ is a section of
$\ad P\otimes T^*X$ while $\alpha^\perp$ is a section of $(\ad
P)^\perp\otimes T^*X$ with $(\ad P)^\perp=\ad Q/\ad P$. Furthermore
$\coker\sigma_P\cong(\ad P)^\perp\otimes T^*X$ and given the
isomorphism $\sigma_Q$, we see that $T_0(P)$ can be identified with
$\alpha^\perp$. Equivalently, since by definition
$\nabla\Xi=(\nabla^g+\alpha^\perp)\Xi=0$ for any $\Xi\in\eta$, we have
\begin{equation}
   \text{$T_0(P)$ can be identified with the set
   $\{\nabla^g\Xi:\Xi\in\eta\}$.}
\end{equation}
Finally, if $T_0(P)=0$ then $\nabla^g$ is compatible with $P$ and $X$
has \emph{special holonomy}, that is, for $G\subset O(n)$,
\begin{equation}
   T_0(P)=0 \quad \Leftrightarrow \quad
      \Hol(X)\subseteq G ,
\end{equation}
where $\Hol(X)\equiv\Hol(TX,\nabla^g)$.

A number of examples of such $G$-structures, familiar from the
discussion of special holonomy manifolds, are listed in
table~\ref{tab:Gstructure}. Except for $g$ in $\Spin(7)$ all the
elements of $\eta$ are forms, where, in the table, the subscript
denotes the degree. Consider for instance the case $G=\SU(k)$ in
dimension $n=2k$. This includes Calabi-Yau $k$-folds in the
special case that $T_0(P)=0$. The elements of $\eta$ are the
fundamental two-form $J$ and the complex $k$-form $\Omega$. The
structure $P$ is defined as the set of frames where $J$ and
$\Omega$ have the form
\begin{equation}
\begin{aligned}
   J &= e^1\wedge e^2 + \dots + e^{n-1}\wedge e^n , \\
   \Omega &= (e^1+\mi e^2)\wedge\dots\wedge(e^{n-1}+\mi e^n) .
\end{aligned}
\end{equation}
The two-form $J$ is invariant under
$\Symp(k,\bbR)\subset\GL(2k,\bbR)$ and $\Omega$ is invariant under
$\SL(k,\bbC)\subset\GL(2k,\bbR)$. The common subgroup is
$\SU(k)\subset\SO(2k)$. Thus the pair $J$ and $\Omega$ determine a
metric. For $\SU(k)$-holonomy we then require that the intrinsic
torsion vanishes or equivalently $\nabla^gJ=\nabla^g\Omega=0$ and
$J$ is then the K\"ahler form and $\Omega$ the holomorphic
$k$-form. By considering the corresponding
$\SU(k)$-representations, it is easy to show~\cite{int-tor} that
\begin{equation}
   \text{$T_0(P)$ can be identified with the set
      $\{\dd J, \dd\Omega\}$},
\end{equation}
so that $\Hol(X)\subseteq\SU(k)$ is equivalent to $\{\dd J=0,
\dd\Omega=0\}$. This result that $T_0(P)$ is encoded in the
exterior derivatives $\dd\Xi$ for $\Xi\in\eta$ is characteristic
of all the examples in table~\ref{tab:Gstructure}.
\begin{table}[ht]
\begin{center}
\begin{tabular}{|lllll|}
   \hline
   dimension & special holo. space $X$
      & $G\subset\SO(n)$
      & $\eta$ &  no. of supersyms. \\
   \hline
   $n=2k$ & Calabi--Yau & $\SU(k)$ & $\{J_2 ,\Omega_k\}$
      & $d_\epsilon/2^{k-1}$ \\
   $n=4k$ & hyper-K\"ahler & $\Symp(k)$
      & $\{J_2^{(1)},J_2^{(2)},J_2^{(3)}\}$
      & $d_\epsilon/2^k$ \\
   $n=7$ & $G_2$ & $G_2$ & $\{\phi_3\}$ & $d_\epsilon/8$ \\
   $n=8$ & $\Spin(7)$ & $\Spin(7)$ & $\{g,\Psi_4\}$ & $d_\epsilon/16$  \\
   \hline
\end{tabular}
\end{center}
\caption{$G$-structures and supersymmetry}
\label{tab:Gstructure}
\end{table}
%

%%%%%%%%%%%%%%%%%%%%%%%%%%%%%%%%%%%%%%%%%%%%%%%%%%%%%%%%%%%%%%%%%%%%%%%%%%

\subsection{Supersymmetry and $G$-structures}
\label{sec:flux}

We can now use the language of $G$-structures to characterise the
constraints on the geometry of $X$ due to the existence of solutions
to the Killing spinor equations~\eqref{KS}. We define the space of
solutions
\begin{equation}
\label{eq:CS}
   C = \{ \xi\in \Csec(S^X): D^X\xi=0,\  Q^X\xi=0 \} ,
\end{equation}
which defines a sub-bundle of $S^X$. The basic idea is that the
existence of $C$ implies that there is a sub-bundle $P$ of the frame
bundle and hence a $G$-structure.

First note that since we have spinors we have an $\SO(n)$-structure
$Q$ defined by the metric and orientation and a spin structure
$(\tilde{Q},\pi)$, where $\tilde{Q}$ is a $\Spin(n)$ principle bundle
and $\pi:\tilde{Q}\to Q$ is the covering map modelled on the double
cover $\Spin(n)\to\SO(n)$. For any $n$ the Clifford algebra
$\Cliff(n)$ is equivalent to a general linear group acting on the
vector space of spinors $\xi$ and implying we can also define a
$\Cliff(n)$ principle bundle $\tilde{\Gamma}$ with
$\tilde{Q}\subset\tilde{\Gamma}$. Recall that $D^X$ is a Clifford
connection defined on $\tilde{\Gamma}$ and generically does not
descend to a connection on $\tilde{Q}$.

Let $K_x\subset\Cliff(n)$ be the stabilizer group in the Clifford
algebra of $\rest{C}_x$, the set of solutions $C$ evaluated at the
point $x\in X$. Since $D^X$ is a $\Cliff(n)$ connection, by
parallel transport $K=K_x$ is independent of $x\in X$, and hence
$C$ defines a $K$ principle sub-bundle
$\tilde{\Lambda}\subset\tilde{\Gamma}$ built from  those elements
of $\tilde{\Gamma}$ leaving $C$ invariant. We can equally well
consider the stabilizer $\tilde{G}_x\subset\Spin(n)$ of
$\rest{C}_x$ in the spin group. Since $D^X$ does not descend to
$\tilde{Q}$ in general $\tilde{G}_x$ is not independent of $x\in
X$ and hence the stabilizer does not define a sub-bundle of
$\tilde{Q}$. However, since there is only a finite number of
possible stabiliser groups, we can still define a unique
$\tilde{G}=\tilde{G}_x$ with  $x\in U$ for some open subset of
$U\subset X$ (with possibly non-trivial topology). Or
alternatively we can restrict our considerations to $C$ such that
$\tilde{G}$ is globally defined. In this way $C$ defines a
sub-bundle $\tilde{P}\subset\tilde{Q}$ of the spin bundle. The
double cover $\pi$ then restricts to a projection
$\pi:\tilde{P}\to P\subset Q$ and hence we have a $G$-structure
$P$ where $G$ is the projection of $\tilde{G}$. (In fact in all
cases we consider $\tilde{G}=G$ and $P\cong\tilde{P}$.) In
conclusion we see that
\begin{quote}
\begin{itemize}
\item[(1)] $C$ defines a $G$-structure $P$ over (at least) some open
   subset $U\subset X$ where $G\subset\SO(n)$.
\end{itemize}
\end{quote}
The different structures and groups can be summarized as follows
\begin{equation}
\label{structures}
   \begin{CD}
      \tilde{\Gamma} @<<< \tilde{\Lambda} \\
      @AAA @AAA \\
      \tilde{Q} @<<< \tilde{P} \\
      \pi@VVV \pi@VVV \\
      Q @<<< P
   \end{CD}
   \qquad \qquad
   \begin{CD}
      \Cliff(n) @<<< K \\
      @AAA @AAA \\
      \Spin(n) @<<< \tilde{G} \\
      \pi@VVV \pi@VVV \\
      \SO(n) @<<< G
   \end{CD}
\end{equation}
Note we can equivalently think of defining $\tilde{P}$ as the
intersection $\tilde\Lambda\cap\tilde{Q}$ as embeddings in
$\tilde{\Gamma}$. Generically this is not a bundle defined over
the whole of $X$ since the fibre group can change, reflecting the
fact that $P$ is generically only defined over $U\subset X$. Note
also that, by construction,
\begin{equation}
\label{eq:holDX}
   \Hol(S^X,D^X) \subseteq K .
\end{equation}
This corresponds to the notion of generalised holonomy introduced
by Duff and Liu~\cite{genhol}. Note, however, that this misses the
important information that there is also a spin structure
$\tilde{Q}\subset\tilde{\Gamma}$ in the Clifford bundle. In other
words the full information is contained in the pair
$(\tilde{\Lambda},\tilde{Q})$, which at least in a patch $U$
translates into the $G$-structure $P$.

To see explicitly that $C$ defines a $G$-structure recall that the
Clifford algebra gives us a set of maps $w_p:\Csec(S^X\otimes
S^X)\to\Csec(\Lambda^pTX)$ given by
\begin{equation}
\label{eq:iso}
   (\xi,\chi) \mapsto
      w_p(\xi,\chi) = \bar{\xi}\gamma^{(p)}\chi ,
\end{equation}
where $\gamma^{(p)}$ is the antisymmetric product of $p$ gamma
matrices generating the Clifford algebra $\Cliff(n)$. By construction
if $\xi,\chi\in C$ then $w_p(\xi,\chi)$ is invariant under $G$. The
invariant forms $\Xi\in\eta$ defining $P$ are then generically
constructed from combinations of bilinears of the form
$w_p(\xi,\chi)$. Specific examples will be given in the next section.

Finally, since $D^X$ is determined by the flux, dilaton, and
$\Delta$ or $\lambda$ and $m$, from the discussion of the last
section, we have our second result
\begin{quote}
\begin{itemize}
\item[(2)] the intrinsic torsion $T_0(P)$ is determined in terms of the flux,
   dilaton, and $\Delta$ or $\lambda$ and $m$.
\end{itemize}
\end{quote}
Generically, however, there may be components of, for instance, the
flux which are not related to $T_0(P)$. Thus we see that the
existence of solutions to the Killing spinor equations~\eqref{KS}
translates into the existence of a $G$-structure $P$ with specific
intrinsic torsion $T_0(P)$.

As mentioned above, in some cases the $G$-structure is globally
defined. On the other hand, in some cases the $G$-structure is
only defined locally in some open set, and possibly only in a
topologically trivial neighbourhood. Of course in such a
neighbourhood the structure group of the frame bundle can always
be reduced to the identity structure. However, the key point is
that supersymmetry defines a canonical $G$-structure that can be
used to give a precise characterisation of the local geometry of
the solution. In turn, as we shall see, this provides an often
powerful method to construct explicit local supersymmetric
solutions. Furthermore, the global properties of such solutions
can then be found by determining the maximal analytic extension of
the local solution (this is a standard technique used in the
physics literature).

Let us now see how this description in terms of $G$-structures
works in a couple of specific examples relevant to the new
solutions we will discuss later.

\subsubsection*{Example 1: $n=6$ in Type IIB}

Consider the case of type IIB supergravity with
$M=\bbR^{3,1}\times X$ and only the self-dual five-form
non-vanishing (the most general case is considered in
ref.~\cite{Dall'Agata:2004dk}). First we need the spinor
decomposition. Recall that
$\epsilon=\left(\begin{smallmatrix}\epsilon_1\\ \epsilon_2
   \end{smallmatrix}\right)$ is a section of $S=S_+\oplus S_+$ where
the spin bundle $S_+$ corresponds to the real (positive) chirality
spinor representation $\Delta^{+\bbR}_{9,1}$ of $\Spin(9,1)$. In
general we have that the complexified representation decomposes as
\begin{equation}
   \left(\Delta^{+\bbR}_{9,1}\right)_\bbC =
      \Delta^+_{3,1}\otimes\Delta^+_6 +
      \overline{\Delta^+_{3,1}}\otimes\overline{\Delta^+_6} ,
\end{equation}
where $\Delta^+_{3,1}$ and $\Delta^+_6$ are the complex positive
chirality representations of $\Spin(3,1)$ and $\Spin(6)$. The bar
denotes the conjugate representation. Let $S^+_{3,1}$ and $S^+_6$ be
the corresponding spin bundles. To ensure that the $\epsilon_i$
are real we decompose
\begin{equation}
   \epsilon_i =
      \psi \otimes \me^{\Delta/2}\xi_i +
      \psi^c \otimes \me^{\Delta/2}\xi_i^c ,
\end{equation}
where $\psi\in\Csec(S^+_{3,1})$, $\xi_i\in\Csec(S^+_6)$ while $\psi^c$
and $\xi_i^c$ are the complex conjugate spinors and we have included
factors of $\me^{\Delta/2}$ in the definition of $\xi_i$ for convenience.
We then define the combinations $\xi^\pm=\xi_1\pm\mi\xi_2$.

The self-dual five-form flux ansatz~\eqref{flux-ansatz} can be written
as
\begin{equation}
   F^{(5)} = \me^{4\Delta}\vol_{\eta_4}\wedge f - *_X f ,
     \qquad f \in \Csec(\Lambda^1T^*X) ,
\end{equation}
where $*_X$ is the Hodge star defined using $g_X$ on $X$.
Decomposing the Killing spinor equations it is easy to show that
either $\xi^+=0$ or $\xi^-=0$. Let us assume that $\xi^-=0$ then,
defining $\xi=\xi^+$ with $S^X=S^+_6$, we have
\begin{equation}
\label{IIBKS}
\begin{aligned}
   D^X &= \nabla^{g_X} + \frac{1}{8}f \lrcorner \gamma^{(2)} , \\
   Q^X &= \gamma^{(1)}\lrcorner \dd\Delta
      + \frac{1}{4}\gamma^{(1)}\lrcorner f .
\end{aligned}
\end{equation}
Note that $D^X$ involves only $\gamma^{(2)}$ and so in this case
it does descend to a metric compatible connection $\nabla$ on
$TX$. Thus, in this case, the $G$-structure to be discussed next,
is in fact globally defined.

We will consider the case of the minimum number of preserved
supersymmetries where the set of solutions $C$ is one-dimensional,
corresponding to non-zero multiples of some fixed solution $\xi\in
C$. The stabiliser of a single spinor is $\SU(3)$ and thus we have
\begin{equation}
   \text{$X$ has $\SU(3)$-structure} .
\end{equation}
It is easy to show that $\nabla^{g_X}(\bar{\xi}\xi)=0$. If
we choose to normalise such that $\bar{\xi}\xi=1$, it then follows
that the elements of $\eta$ fixing the $\SU(3)$ structure are given by
the bilinears 
\begin{equation}
   J = - \mi \bar{\xi}\gamma^{(2)}\xi , \qquad
   \Omega = \bar{\xi}^c\gamma^{(3)}\xi .
\end{equation}

We next calculate the intrinsic torsion. Recall that this is contained
in $\dd J$ and $\dd\Omega$. From~\eqref{IIBKS} one finds
\begin{equation}
\label{IIBit}
\begin{aligned}
   \dd(\me^{4\Delta}) &= - \me^{4\Delta}f , \\
   \dd(\me^{2\Delta}J) &= 0 , \\
   \dd(\me^{3\Delta}\Omega) &= 0 ,
\end{aligned}
\end{equation}
which completely determines the intrinsic torsion, as well as the
flux, in terms of $\dd\Delta$. This implies that $g_X$ is conformally
Calabi--Yau, that is we can write
\begin{equation}
\label{confCY}
   g_X = \me^{-2\Delta}g_6 ,
\end{equation}
where $g_6$ has integrable $\SU(3)$-structure. In addition
\begin{equation}
\label{fexp}
   f = - 4 \dd\Delta ,
\end{equation}

Note that the Bianchi identity for $F^{(5)}$ is satisfied provided we
have $\dd *_X f=0$. This translates in to the harmonic condition
\begin{equation}
\label{laplace}
   \nabla^2_{g_6} \me^{-4\Delta} = 0 ,
\end{equation}
where $\nabla^2_{g_6}$ is the Laplacian for $g_6$ on $X$. Thus we have
completely translated the conditions for a supersymmetric background
into a geometrical constraint~\eqref{confCY} together with a solution
of the Laplacian~\eqref{laplace}.

\subsubsection*{Example 2: $n=7$ in $d=11$}

Now consider the case of $d=11$ supergravity on
$M=\bbR^{3,1}\times X$. (Here we are following the discussion of
ref.~\cite{kmt}.) Again we start with the spinor decomposition. Recall 
that the $d=11$ spinor $\epsilon$ is a section of a spin bundle
corresponding to the real 32-dimensional representation
$\Delta^\bbR_{10,1}$ of $\Spin(10,1)$. Under
$\Spin(3,1)\times\Spin(7)$ the complexified representation decomposes
as 
\begin{equation}
   \big(\Delta^{\bbR}_{10,1}\big)_\bbC =
      \Delta^+_{3,1}\otimes\big(\Delta^\bbR_7\big)_\bbC +
      \overline{\Delta^+_{3,1}}\otimes\big(\Delta^\bbR_7\big)_\bbC ,
\end{equation}
where $\Delta^+_{3,1}$ and $\Delta^\bbR_7$ are the complex positive
chirality representation of $\Spin(3,1)$ and real representation of
$\Spin(7)$ respectively. Let $S^+_{3,1}$ and $S^\bbR_7$ be the
corresponding spin bundles. To ensure that the $\epsilon_i$
are real we decompose
\begin{equation}
   \epsilon =
      \psi \otimes \left(\xi_1+\mi\xi_2\right) +
      \psi^c \otimes \left(\xi_1-\mi\xi_2\right) ,
\end{equation}
where $\psi\in\Csec(S^+_{3,1})$, $\xi_i\in\Csec(S^\bbR_7)$.

In the flux ansatz~\eqref{flux-ansatz} we assume $G$ is pure magnetic
so
\begin{equation}
   G \in \Csec(\Lambda^4T^*X) .
\end{equation}
Defining $\xi$ as the doublet
$\xi=\left(\begin{smallmatrix}\xi_1\\ \xi_2\end{smallmatrix}\right)$
with $S^X=S^\bbR_7\oplus S^\bbR_7$, we find
\begin{equation}
\label{MKS}
\begin{aligned}
   D^X &= \nabla^{g_X} \otimes \id
      + \frac{1}{12}\gamma^{(2)}\lrcorner {*_X G} \otimes \mi\sigma_2
      + \frac{1}{6}\mi\gamma^{(3)}\lrcorner G \otimes \mi\sigma_2 , \\
   Q^X &= \gamma^{(1)}\lrcorner \dd\Delta \otimes \id
      + \frac{1}{6}\mi\gamma^{(4)}\lrcorner G \otimes \mi\sigma_2 ,
\end{aligned}
\end{equation}
where
$\mi\sigma_2=\left(\begin{smallmatrix}0&1\\-1&0\end{smallmatrix}\right)$.
Note that $D^X$ does not descend to a metric compatible connection
$\nabla$ on $TX$.

Again we are interested in the minimum number of preserved
supersymmetries so the set of solutions $C$ is one-dimensional,
corresponding to non-zero multiples of some fixed solution $\xi\in
C$. In addition we will assume the $\xi_i$ in $\xi$ are each non-zero
and more importantly
\begin{equation}
\label{orthog}
   \bar{\xi_1}\xi_2 = 0 .
\end{equation}
(Note that the generic conditions, without this assumption, were
derived in ref.~\cite{Lukas:2004ip}.)
It is then easy to show that the $\me^{-\Delta/2}\xi_i$ have constant
norm. Together with~\eqref{orthog} this then implies that the
stabliser of
$\xi=\left(\begin{smallmatrix}\xi_1\\ \xi_2\end{smallmatrix}\right)$
is $G=\SU(3)$ independent of $x\in X$ and hence
\begin{equation}
   \text{$X$ has $\SU(3)$-structure.}
\end{equation}
Note that this is an $\SU(3)$-structure in seven dimensions.

If we normalise $\xi\in C$ such that
$\me^{-\Delta}\bar{\xi}_1\xi_1=\me^{-\Delta}\bar{\xi}_2\xi_2=1$ then
the elements of $\eta=\{J,\Omega,K\}$ fixing the $\SU(3)$ structure
are given by
\begin{equation}
\label{SU3sol}
\begin{aligned}
   J &= - \me^{-\Delta} \bar{\xi}_1\gamma^{(2)}\xi_2 , \\
   \Omega &= -\tfrac{1}{2}\me^{-\Delta}\left(
         \bar{\xi}_1\gamma^{(3)}\xi_1 - \bar{\xi}_2\gamma^{(3)}\xi_2
         \right)
      + \mi \me^{-\Delta}\bar{\xi}_1\gamma^{(3)}\xi_2 , \\
   K &= - \mi \me^{-\Delta}\bar{\xi}_1\gamma^{(1)}\xi_2 ,
\end{aligned}
\end{equation}
where we are using the convention $\mi\gamma^{(7)}=\vol_X\id$. The
one-form $K$ defines a product structure $R\subset Q$ with fibre
$\SO(6)\subset\SO(7)$ and then $J$ and $\Omega$ define the
$G$-structure $P\subset R$ with fibre $\SU(3)$.

As in six dimensions the intrinsic torsion of an $\SU(3)$-structure in
seven dimensions is completely determined by the exterior derivatives
of $K$, $J$ and $\Omega$. One finds
\begin{equation}
\label{Mit}
\begin{aligned}
   \dd (\me^{2\Delta} K) &= 0 , \\
   \dd (\me^{4\Delta} J) &= \me^{-4\Delta}*_X G , \\
   \dd(\me^{3\Delta}\Omega) &= 0 , \\
   \dd (\me^{2\Delta}J\wedge J) &= - 2\,\me^{2\Delta} G \wedge K .
\end{aligned}
\end{equation}
These equations were derived in ref.~\cite{kmt} (up to a factor in the 
last equation differs, as discussed in ref.~\cite{m6}). It was argued
in ref.~\cite{Dall'Agata:2003ir} that these are the necessary and
sufficient conditions for a geometry to admit a single Killing
spinor. Furthermore, the second equation implies the $G$ equation of
motion and thus, given an integrability argument as in
ref.~\cite{11d}, only the Bianchi identity $\dd G$ need be imposed to
give a solution to the full equations of motion. 

%%%%%%%%%%%%%%%%%%%%%%%%%%%%%%%%%%%%%%%%%%%%%%%%%%%%%%%%%%%%%%%%%%%%%%%%%%

\subsection{Relation to generalised calibrations}
\label{sec:calib}

In turns out that there is a very interesting relation between the
torsion conditions, such as~\eqref{IIBit} and~\eqref{Mit}, one
derives for supersymmetric backgrounds and the notion of a
``generalised calibrations'' introduced in ref.~\cite{g-cal}. This
gives a very physical interpretation of the conditions in terms of
string theory ``branes''. Here we will only touch on this relation
briefly.

Let us start by recalling the notion of calibrations and a
calibrated cycle~\cite{calib,sp-calib} (for a review see
refs.~\cite{joyce,jerome}) Suppose we have a Riemannian
manifold $X$ with metric $g_X$ and let $\xi\subset T_xX$ be an
oriented $p$-dimensional tangent plane at any point $x\in X$. We
can then define $\vol_\xi$ as the volume form on $\xi$ built from
the restriction $\rest{g_X}_\xi$ of the metric to $\xi$. A
$p$-form $\Xi$ is then a \emph{calibration} if
\begin{equation}
\begin{aligned}
   \text{(i)} & \qquad &
      \rest{\Xi}_\xi &\leq \vol_\xi && \forall \; \xi , \\
   \text{(ii)} & \qquad &
      \dd \Xi &= 0 .
\end{aligned}
\label{calib}
\end{equation}
Furthermore, given a $p$-dimensional oriented submanifold $C_p$, we
say $C_p$ is \emph{calibrated} if
\begin{equation}
\label{eq:cal-C}
   \text{calibrated submanifold:} \quad \rest{\Xi}_{T_xC_p}
      =  \vol_{T_xC_p} \quad \forall \, x \in C_p .
\end{equation}
If $C'_p$ is another submanifold in the same homology class we have
\begin{equation}
   \int_{C_p}\vol_{C_p} = \int_{C_p}\rest{\Xi}_{TC_p}
      = \int_{C'_p}\rest{\Xi}_{TC'_p} \leq \int_{C'_p}\vol_{C'_p} ,
\end{equation}
and we get the main result that a calibrated submanifold has minimum
volume in its homology class.

Now suppose we have a set of invariant tensors $\eta$ defining a
$G$-structure $P$ with $G\subset\SO(n)$. One finds that, for $p$-forms
$\Xi$
\begin{equation}
\begin{aligned}
   \Xi \in \eta \quad &\Rightarrow \quad
      \text{calibration condition~(i)} , \\
   T_0(P)=0 \quad &\Rightarrow \quad
      \text{calibration condition~(ii)} . \\
\end{aligned}
\end{equation}
Thus the vanishing intrinsic torsion of the $G$-structure (i.e.
special holonomy $G$) corresponds to the closure of the
calibration forms. It is natural, then, to try to interpret our
intrinsic torsion conditions~\eqref{IIBit} and~\eqref{Mit} as
defining ``generalised calibrations''~\cite{g-cal}. Obviously
calibrated sub-manifolds will no longer be volume minimising, but
one might ask if there is some more general notion of the
``energy'' of the submanifold which is minimised by calibrated
sub-manifolds.

String theory provides precisely such an interpretation. It
contains a number of extended $p+1$-dimensional objects which
embed into the spacetime and are known as ``$p$-branes'': a simple
example is the two-dimensional string itself. Each brane has a
particular energy functional depending both on the volume of the
embedded submanifold and crucially the flux and dilaton.
Differential conditions such as~\eqref{IIBit} and~\eqref{Mit} then
imply that the corresponding brane energy is minimised when the
submanifold is calibrated by a generalised calibration.

Consider for instance our $d=11$ example with $M=\bbR^{3,1}\times
X$. The relevant branes in eleven dimensions are the ``M2-brane'' and
the ``M5-brane'' and are described by embeddings of the worldvolumes
$\Sigma\hookrightarrow M$. In particular, we can take
$\Sigma=\bbR^{r,1}\times C_s$ with $r+s\in\{2,5\}$, where $C_s$ is a
$p$-dimensional submanifold of $X$. One then says that the brane is
``wrapped'' on $C_s$. Each of the conditions~\eqref{Mit} can then be
interpreted as generalised calibrations for different types of wrapped
brane. We have
\begin{equation}
\begin{aligned}\label{gcalexps}
   \dd (\me^{2\Delta} K) &= 0
      &\quad& \text{M2-brane on $C_1$} , \\
   \dd (\me^{4\Delta} J) &= \me^{-4\Delta}*_X G
      &\quad& \text{M5-brane on $C_2$} , \\
   \dd(\me^{3\Delta}\Omega) &= 0
      &\quad& \text{M5-brane on $C_3$} , \\
   \dd (\me^{2\Delta}J\wedge J) &= - 2\,\me^{2\Delta} G \wedge K
      &\quad& \text{M5-brane on  $C_4$} .
\end{aligned}
\end{equation}
Note that the power of $\me^\Delta$ appearing in each expression
counts the $q+1$ unwrapped dimensions of the brane. Roughly, the
fluxes appearing on the right hand side can be understood by
noting that M2-branes couple to electric $G$-flux, that M5-branes
couple to magnetic $G$-flux and that we have only kept certain
components of $G$ in our ansatz (for example the electric $G$-flux
vanishes). The flux appearing in the last expression in
\eqref{gcalexps} arises from the fact that there can be induced
M2-brane charge on the M5-brane. For more on this correspondence
see refs.~\cite{Gauntlett:2001ur,11d,int-tor,Martelli:2003ki}.

%%%%%%%%%%%%%%%%%%%%%%%%%%%%%%%%%%%%%%%%%%%%%%%%%%%%%%%%%%%%%%%%%%%%%%%%%%

\section{New $AdS_5$ solutions in M-theory}
\label{sec:M}

We now turn to the specific problem of finding, first, the generic
structure of the minimal supersymmetric configurations of $D=11$
supergravity with $M=\AdS_5\times X$ and, second, a class of
particular solutions of this form. Such backgrounds are of
particular interest because, via the AdS-CFT correspondence, they
are dual to $\mathcal{N}=1$ superconformal field theories in
four-dimensions. This work was first presented in ref.~\cite{m6}.

%%%%%%%%%%%%%%%%%%%%%%%%%%%%%%%%%%%%%%%%%%%%%%%%%%%%%%%%%%%%%%%%%%%%%%%%%%

\subsection{General differential conditions}
\label{sec:Mconds}

As we have seen $\AdS_5\times X$ geometries are special cases of
$\bbR^{3,1}\times X'$ where $X'=X\times\bbR$ with metrics and warp
factors related as in eqns.~\eqref{reduction}. Thus we can
actually obtain the general conditions for supersymmetric $\AdS_5$
compactifications from the corresponding $n=7$ $\SU(3)$ conditions
given in eq.~\eqref{Mit}. (One might be concerned that these
latter conditions are not completely generic, nonetheless one can
show~\cite{m6} that they give the generic conditions for
$\AdS_5$.)

To derive the conditions explicitly, let us denote the $\SU(3)$
structure on $X'$ by the primed forms $(K',J',\Omega')$. The radial
unit one-form $\me^\lambda\dd r$ is generically not parallel to
$K'$; instead we can write
\begin{equation}
   \me^\lambda\dd r = -\sin\zeta K' - \cos\zeta W' , \\
\end{equation}
where $W'$ is a unit one-form orthogonal to $K'$. We can then define
two other unit mutually orthogonal one-forms
\begin{equation}
\label{K1def}
\begin{aligned}
   K^1 &= \cos\zeta K' - \sin\zeta W' \\
   K^2 &= V' = J'\cdot W'
\end{aligned}
\end{equation}
where $K^1$ is the orthogonal linear combination of $K'$ and $W'$ and
$K^2$ is defined using $J'_a{}^b$ the almost complex structure on
$X'$. We can then define real and imaginary two-forms from the parts
of $J$ and $\Omega$ orthogonal to $W'$ and $V'$, that is
\begin{equation}
\label{SU2}
\begin{aligned}
   J &= J' - W' \wedge V' \\
   \Omega &= i_{W'+\mi V'}\Omega'~.
\end{aligned}
\end{equation}
Note that $\Omega$ is not strictly a two-form on $X$ but is a
section of $\Lambda^2T^*X$ twisted by the complex line bundle defined
by $W'+\mi V'$. This implies that the set $(K^1,K^2,J,\Omega)$
actually defines a local $U(2)$ structure on the six-dimensional
manifold $X$, rather than an $\SU(2)$ structure as would be the case if
$\Omega$ were truly a two-form. Note that the structure is only local
since, in particular, it breaks down when $K'$ is parallel to $\dd r$,
that is $\cos\zeta=0$, in which case we cannot define $K^1$ and
$K^2$. Using these definitions the constraints~\eqref{Mit} become
\begin{align}
   \dd(\me^{3\lambda}\sin\zeta)
      &= 2 m\,\me^{2\lambda}\cos\zeta\, K^1 , \label{ns.a}\\
   \dd (\me^{4\lambda}\cos\zeta\, \Omega)
      &= 3 m\,\me^{3\lambda} \Omega \wedge (- \sin\zeta K^1 + i  K^2) ,
      \label{ns.c}\\
   \dd(\me^{5\lambda}\cos\zeta\, K^2)
      &= \me^{5\lambda}* G
         + 4 m\,\me^{4\lambda}(J - \sin\zeta K^1\wedge K^2) ,
         \label{ns.d}\\
   \dd (\me^{3\lambda}\cos\zeta\, J \wedge K^2)
      &= \me^{3\lambda}\sin\zeta G
         + m\,\me^{2\lambda} (J \wedge J
            - 2 \sin\zeta J \wedge K^1 \wedge K^2 ) . \label{ns.e}
\end{align}
(Note that the $\SU(2)$ structure here differs from that used in
ref.~\cite{m6} by a conformal rescaling.)

To ensure we have a solution of the equations of motion, in
general one also needs to impose the equation of motion and
Bianchi identity for $G$. The connection with the $n=7$ results gives
us a quick way of seeing that, in fact, provided $\sin\zeta$ is not
identically zero, both conditions are a consequence of the supersymmetry
constraints~\eqref{ns.a}--\eqref{ns.e}. As already noted, the equation
of motion for $G$ follows directly from the exterior
derivative of the second equation in~\eqref{Mit}. For the
Bianchi identity one notes that, given the ansatz for the $n=7$
metric and $G$, the first and last equations in~\eqref{Mit}
imply in general that
\begin{equation}
   \sin\zeta \dd G \wedge \dd r = 0
\end{equation}
since $\dd G$ lies solely in $X$. This implies that $\dd G=0$
provided $\sin\zeta$ is not identically zero -- which can only
occur only when $m=0$ (from ~\eqref{ns.a}). Thus we see that the
constraints~\eqref{ns.a}--\eqref{ns.e} are necessary and
sufficient both for supersymmetry and for a solution of the
equations of motion.

%%%%%%%%%%%%%%%%%%%%%%%%%%%%%%%%%%%%%%%%%%%%%%%%%%%%%%%%%%%%%%%%%%%%%%%%%%

\subsection{Local form of the metric}
\label{sec:metric}

By analysing the differential
conditions~\eqref{ns.a}--\eqref{ns.e} on the forms, after some
considerable work, one can derive the necessary and sufficient
conditions on the local form of the metric and flux. Here we will
simply summarize the results referring to ref.~\cite{m6} for more
details.

First one notes that as a vector $e^{-\lambda}\cos\zeta\, K_2$ is
Killing and that coordinates can be chosen so that
\begin{equation}
\label{eq:Killing}
   \frac{\del}{\del\psi} = \frac{1}{3m}\me^{-\lambda}\cos\zeta\, K_2
\end{equation}
In addition the Lie derivatives
$\mathcal{L}_{\del/\del\psi}G=\mathcal{L}_{\del/\del\psi}\lambda=0$
vanish so in fact acting with $\mathcal{L}_{\del/\del\psi}$
preserves the full solution. This reflects the fact that the dual
field theory has a $U(1)_R$ symmetry.

Second, one can introduce a coordinate $y$ for $K_1$ given by
\begin{equation}
   2m y = \me^{3\lambda} \sin\zeta
\end{equation}
so that
\begin{equation}
\label{K1sol}
   K^1 = \me^{-2\lambda}\sec\zeta \dd y~.
\end{equation}
While we could eliminate either $\lambda$ or $\zeta$ from the following
formulae, for the moment it will be more convenient to keep both.

The metric then takes the form
\begin{equation}
\label{g6}
   g_X  =  \me^{-4\lambda}\left( \hat{g}
         + \sec^2\zeta \dd y\otimes \dd y \right)
      + \frac{1}{9m^2}\me^{2\lambda} \cos^2\zeta
         (\dd\psi + \rho)\otimes(\dd\psi + \rho)
\end{equation}
where $i_{\partial_y}\rho=i_{\partial_\psi}\rho=0$. We also have,
with $\hat{J}=\me^{4\lambda}J$,
\begin{align}
   \text{(a)} & \quad\text{$\del/\del\psi$ is a Killing vector}
      \label{killingvec} \\
   \text{(b)} & \quad\text{$\hat{g}$ is a family of K\"ahler metrics
     on $M_4$ parameterized by $y$}
      \label{kahler} \\
   \text{(c)} & \quad\text{the corresponding complex structure
     $\hat{J}_i{}^j$ is independent of $y$ and $\psi$.}
      \label{integ}
\end{align}
and
\begin{align}
   \label{ydef}
   && && && && \text{(d)} &&
      2m y &= \me^{3\lambda} \sin\zeta && && && && \\
   \label{rhodef}
   && && && && \text{(e)} &&
      \rho &= \hat{P} + \hat{J}\cdot\dd_4\log\cos\zeta
\intertext{
where, in complex co-ordinates,
$\hat{P}=\frac{1}{2}\hat{J}\cdot\dd\log\sqrt{\hat{g}}$ is the
canonical connection defined by the K\"ahler metric, satisfying
$\hat{\Re}=\dd\hat{P}$ where $\hat{\Re}$ is the Ricci form. Finally we
have the conditions}
   \label{dyJ}
   && && && && \text{(f)}&&
      \del_y \hat{J} &= - \frac{2}{3} y \dd_4 \rho \\
   \label{dyg}
   && && && && \text{(g)}&&
      \del_y\log\sqrt{\det\hat{g}} &=
         -3y^{-1}\tan^2\zeta - 2\del_y\log\cos\zeta~.
\end{align}
The four-form flux $G$ is given by
\begin{equation}
\label{genflux}
\begin{aligned}
   G &= -(\partial_y \me^{-6\lambda})\widehat{\rm{vol}}_4
      - \me^{-10\lambda}\sec\zeta
         (\hat *_4\dd_4 \me^{6\lambda})\wedge K^1
      - \frac{1}{3m}\me^{-\lambda}\cos^3\zeta
         (\hat *_4\partial_y \rho)\wedge K^2 \\
      &\qquad
      + \me^{\lambda}\left[
         \frac{1}{3m}\cos^2\zeta \hat*_4\dd_4\rho
           - 4me^{-6\lambda}\hat J \right] \wedge K^1\wedge K^2
\end{aligned}
\end{equation}
and is independent of $\psi$ -- that is,
$\mathcal{L}_{\del/\del\psi}G=0$. As discussed previously the
equations of motion for $G$ and the Bianchi identity are implied by
expressions~\eqref{killingvec}--\eqref{dyg}.

To summarize, we have given the local form of the generic
$\mathcal{N}=1$ $\AdS_5$ compactification in $d=11$ supergravity.
Any $d=11$ AdS-CFT supergravity dual of a $d=4$ superconformal
field theory will have this form.

%%%%%%%%%%%%%%%%%%%%%%%%%%%%%%%%%%%%%%%%%%%%%%%%%%%%%%%%%%%%%%%%%%%%%%%%%%

\subsection{Complex $X$ ansatz}
\label{sec:M-ansatz}

In this section we consider how the conditions on the metric
specialise for solutions where the six-dimensional space $X$ is a
complex manifold. Crucially, the supersymmetry conditions simplify
considerably and we are able to find many solutions in closed
form. Globally, the new regular compact solutions that we
construct are all holomorphic $\CP^1$ bundles over a smooth
four-dimensional K\"ahler base $M_4$. Using a recent mathematical
result on K\"ahler manifolds~\cite{apostolov}, we are able to
classify completely this class of solutions (assuming that the
Goldberg conjecture is true).  In particular, at fixed $y$ the
base is either (i) a K\"ahler--Einstein (KE) space or (ii) a
non-Einstein space which is the product of two constant curvature
Riemann surfaces.

More precisely we specialize to the case where
\begin{equation*}
   \text{$g_X$ is a Hermitian metric on a complex manifold $X$,}
\end{equation*}
where we define the complex structure, compatible with $g_X$ and the
local $U(2)$-structure, given by the holomorphic three-form
$\Omega_{(3)}=\Omega\wedge(K^1+\mi K^2)$. Requiring this complex
structure to be integrable, that is
$\dd\Omega_{(3)}=A\wedge\Omega_{(3)}$ for some $A$, implies that
\begin{equation}
\label{allcondcom}
   \dd_4\zeta = 0 \qquad
   \dd_4\lambda = 0 \qquad
   \del_y\rho = 0 .
\end{equation}
In addition one finds that the connection $\rho$ is simply the
canonical connection defined by the K\"ahler metric $\hat{g}$, that is
\begin{equation}
\label{r=P}
   \rho = \hat{P}
\end{equation}
together with the useful condition that
\begin{equation*}
   \text{at fixed $y$, the Ricci tensor on $\Ricci{\hat{g}}$ has two
   pairs of constant eigenvalues.}
\end{equation*}

We would like to find global regular solutions for the complex
manifold $X$. Our construction is as follows. We require that $\psi$
and $y$ describe a holomorphic $\CP^1$ bundle over a smooth K\"ahler
base $M_4$
\begin{equation}
\label{fibration}
   \begin{CD}
      \CP_{y,\psi}^1 @>>> X \\
      && @VVV \\
      && M_4
   \end{CD}
\end{equation}
For the $(y,\psi)$ coordinates to describe a smooth $\CP^1$
we take the Killing vector $\del/\del\psi$ to have compact orbits so
that $\psi$ defines an azimuthal angle and $y$ is taken to lie in the
range $[y_1,y_2]$ with $\cos\zeta(y_i)=0$. Thus $y_i$ are the two
poles where the $U(1)$ fibre shrinks to zero size. It turns out that
the metric $g_X$ gives a smooth $S^2$ only if we choose the
period of $\psi$ to be $2\pi$. Given the connection~\eqref{r=P}, we
see that, as a complex manifold,
\begin{equation}
\label{topology}
   X = \bbP(\mathcal{O}\oplus\mathcal{L}) ,
\end{equation}
where $\mathcal{L}$ is the canonical bundle and $\mathcal{O}$
the trivial bundle on the base $M_4$.

Let us now consider the K\"ahler base. A recent result on K\"ahler
manifolds (Theorem~2 of ref.~\cite{apostolov}) states that, if the
Goldberg conjecture\footnote{The Goldberg conjecture says that any
compact Einstein almost K\"ahler manifold is K\"ahler-Einstein
i.e. the complex structure is integrable. This has been proven for
non-negative curvature~\cite{Sekigawa}} is true, then a compact
K\"ahler four-manifold whose Ricci tensor has two distinct pairs
of constant eigenvalues is locally the product of two Riemann
surfaces of (distinct) constant curvature. If the eigenvalues are
the same the manifold is by definition K\"ahler--Einstein. The
compactness in the theorem is essential, since there exist
non-compact counterexamples. However, for AdS/CFT purposes, we are
most interested in the compact case (for example, the central
charge of the dual CFT is inversely proportional to the volume).

From now on we will consider only these two cases. One then finds that
the conditions~\eqref{dyJ} and~\eqref{dyg} can be partially
integrated. In summary we have two cases:
\begin{equation}
\begin{aligned}
   \text{case 1:} \qquad \hat{g} &=
      \tfrac{1}{3}\left(b-ky^2\right)\tilde{g}_k
      \\*[0.3cm]
   \text{case 2:} \qquad \hat{g} &=
      \tfrac{1}{3}\left(a_1-k_1y^2\right)\tilde{g}_{k_1}
      + \tfrac{1}{3}\left(a_2-k_2y^2\right)\tilde{g}_{k_2} \\
\end{aligned}
\end{equation}
where $k,k_i\in\{0,\pm1\}$, and the (two- or four-dimensional)
K\"ahler--Einstein metrics $\tilde{g}_k$ satisfy
\begin{equation}
\label{KECOND}
   \Ricci{\tilde{g}_k} = k \tilde{g}_k
\end{equation}
and are independent of $y$. The remaining equation~\eqref{dyg}, implies
\begin{equation}
\label{diffeq}
\begin{aligned}
   \text{case 1:} &\qquad
   m^2 (1+6y\del_y\lambda) = \frac{k}{b-ky^2}
      \,(\me^{6\lambda}-4m^2y^2) ,\\
   \text{case 2:} &\qquad
   m^2 (1+6y\del_y\lambda) =
      \frac{k_2a_1+k_1a_2-2k_1k_2y^2}{2(a_1-k_1y^2)(a_2-k_2y^2)}
      \,(\me^{6\lambda}-4m^2y^2)~.
\end{aligned}
\end{equation}
%

%%%%%%%%%%%%%%%%%%%%%%%%%%%%%%%%%%%%%%%%%%%%%%%%%%%%%%%%%%%%%%%%%%%%%%%%%%

\subsection{New compact solutions}
\label{sec:M-sols}

\subsubsection{Case 1: KE base}

We start by considering the case where the base is K\"ahler--Einstein
(KE). The remaining supersymmetry condition~\eqref{diffeq} can be
integrated explicitly. One finds,
\begin{equation}
\label{warpke}
\begin{aligned}
   \me^{6\lambda} &= \frac{2m^2(b-ky^2)^2}{2kb+cy+2k^2y^2} \\
   \cos^2\zeta &= \frac{b^2-6kby^2-2cy^3-3k^2y^4}{(b-ky^2)^2}
\end{aligned}
\end{equation}
where $c$ is an integration constant. Without loss of generality by an
appropriate rescaling of $y$ we can set $b=1$ and $c\geq 0$.

Assuming $X$ has the topology given by~\eqref{topology}, we find this
leads to a smooth metric at the $y=y_i$ poles of the $\CP^1$
fibres provided we take $\psi$ to have period $2\pi$. One then finds
our first result
\begin{quote}
\begin{slshape}
   for $0\leq c<4$ we have a one-parameter family of completely
   regular, compact, complex metrics $g_X$ with the topology of a
   $\CP^1$ fibration over a positive curvature KE base.
\end{slshape}
\end{quote}
For negative ($k=-1$) and zero ($k=0$) curvature KE metrices $\hat{g}$
there are no regular solutions. Since four-dimensional compact
K\"ahler-Einstein spaces with positive curvature have been
classified~\cite{tian,tianyau}, we have a classification for the above
solutions. In particular, the base space is either $S^2\times S^2$ or
$\CP^2$, or $\CP^2\#_n \CP^2$ with $n=3,\dots,8$. For the first two
examples, the KE metrics are of course explicitly known and this gives
explicit solutions of M-theory. The remaining metrics, although proven
to exist, are not explicitly known, and so the same applies to the
corresponding M-theory solutions.

\subsubsection{Case 2: product base}

Next consider the case where the base is a product of constant
curvature Riemann surfaces. Again the remaining supersymmetry
condition~\eqref{diffeq} can be integrated explicitly giving
\begin{equation}
\begin{aligned}
   \me^{6\lambda} &=
      \frac{2m^2(a_1-k_1y^2)(a_2-k_2y^2)}
         {(k_2a_1+k_1a_2)+cy+2k_1k_2y^2} \\
   \cos^2 \zeta &=
      \frac{a_1a_2-3(k_2a_1+k_1a_2)y^2-2cy^3-3k_1k_2y^4}
         {(a_1-k_1y^2)(a_2-k_2y^2)}
\end{aligned}
\end{equation}
where $c$ is an integration constant giving a three-parameter
family of solutions. Note that on setting $a_1=a_2=b$, $k_1=k_2=k$
these reduce to the KE solutions considered above. Again we have a
smooth metric at the $y=y_i$ poles of the $\CP^2$ fibres provided
we take $\psi$ to have period $2\pi$. The full metric $g_X$ is
regular if the base is $S^2\times T^2$, $S^2\times S^2$ or
$S^2\times H^2$. However the final case is not compact.

Summarizing the compact cases, for $S^2\times T^2$ without loss of
generality we can take $k_2=0$, $a_2=3$ and, by scaling $y$, we can
set $c=1$ or $c=0$. We find
\begin{quote}
\begin{slshape}
   for $0<a<1$ and $c\neq 0$ we have a one-parameter family
   of completely regular, compact, complex metrics $g_X$ where $X$ is
   a topologically trivial $\CP^1$ bundle over $S^2\times T^2$. A
   single additional solution of this type is obtained when $c=0$ and
   $a\neq 0$.
\end{slshape}
\end{quote}
For the $S^2\times S^2$ topology again, generically, one parameter
can be scaled away and we find
\begin{quote}
\begin{slshape}
   for various ranges of $(a_1,a_2,c)$ there are completely regular,
   compact, complex metrics $g_X$ where $X$ is topologically a $\CP^1$
   bundle over $S^2\times S^2$.
\end{slshape}
\end{quote}
In particular there are solutions when $a_1$ is not equal to $a_2$
and hence this gives a broader class of solutions than in the
K\"ahler-Einstein case considered above. The existence of regular
solutions is rather easy to see if one sets $c=0$. Note that we
can also recover the well-known Maldecena--Nu\~{n}ez
solution~\cite{malnun} when the base has topology $S^2\times H^2$,
though the topology is slightly different from the ansatz here.
More details are given in ref.~\cite{m6}.

The $S^2\times T^2$ solutions are of particular interest since
they lead to new type IIA and type IIB supergravity solutions.
Type IIA supergravity arises from $d=11$ supergravity reduced on a
circle. Since these solutions have two Killing directions on the
$T^2$ base we can trivially reduce on one circle in $T^2$ to give
a IIA solution. Given the second Killing vector we can then use
T-duality to generate a IIB solution. (T-duality is a specific map
between IIA and IIB supergravity backgrounds which exists when
each background has a Killing vector which also preserves the flux
and dilaton, and also the Killing spinors if the map is to
preserve supersymmetry at the level of the supergravity solution.)
The resulting IIB background has the form $\AdS_5\times Z$ with
non-trivial $F^{(5)}$ flux. As we will see, this implies that $Z$
is a Sasaki--Einstein manifold. The geometry of these manifolds
will be the subject of the following section.

%%%%%%%%%%%%%%%%%%%%%%%%%%%%%%%%%%%%%%%%%%%%%%%%%%%%%%%%%%%%%%%%%%%%%%%%%%

\section{A new infinite class of Sasaki--Einstein solutions}
\label{sec:SE}

By analogy with the previous section let us now turn to the case
of $\AdS_5\times X$ solutions in IIB supergravity with non-trivial
$F^{(5)}$. It is a well-known result that $X$ must then be
Sasaki--Einstein~\cite{IIBse}. As noted above, the $d=11$
solutions on $S^2\times T^2$ potentially give new $n=5$
Sasaki--Einstein solutions. In this section we discuss the
structure of these solutions. In fact we will show the general
result that
\begin{quote}
   \textsl{for every positive curvature $2n$-dimensional
   K\"ahler--Einstein manifold $B_{2n}$, there is a countably infinite
   class of associated compact, simply-connected, spin,
   Sasaki--Einstein manifolds $X_{2n+3}$ in dimension
   $2n+3$.}
\end{quote}

\subsection{Sasaki--Einstein spaces}

Let us start by showing directly that for IIB backgrounds of the
form $\AdS_5\times X$ with $F^{(5)}$ flux $X$ must be
Sasaki--Einstein. As before we will consider the reduction from
$\bbR^{3,1}\times X'$ to $\AdS_5\times X$ of the backgrounds given
in eqns.~\eqref{IIBit}. Let $(J',\Omega')$ denote the $\SU(3)$
structure on $X'$. Picking out the radial one-form
$R\equiv\me^\lambda\dd r$ globally defines a second one-form
$K=J'\cdot R$. One then has the real and complex two-forms given
by
\begin{equation}
\label{U2}
\begin{aligned}
   J &= J' - K \wedge R \\
   \Omega &= i_{K+\mi R}\Omega'~.
\end{aligned}
\end{equation}
Note that globally $\Omega$ is not strictly a two-form on $X$ but
is a section of $\Lambda^2T^*X$ twisted by the complex line bundle
defined by $K+\mi R$. For this reason $(K,J,\Omega)$ define only a
$U(2)$ (or almost metric contact) structure on $X$ rather than
$\SU(2)$.

Reducing the condition on $(J',\Omega')$ one finds that $\lambda$
is constant and we set it to zero without loss of generality. We
then have that $K$ is unit norm and that
\begin{equation}
\label{SE}
\begin{aligned}
   \dd K &= 2 m J \\
   \dd \Omega &= \mi 3m K \wedge \Omega
\end{aligned}
\end{equation}
with the five-form flux given by
$F^{(5)}=4m(\vol_{AdS_5}+\vol_{X_5})$. Clearly $\mathcal{L}_KJ=0$
and $\mathcal{L}_K\Omega=\mi 3m\Omega$ so that $K$ is a Killing
vector. The second condition in~\eqref{SE} implies that we have an
integrable contact structure. The first condition implies that the metric
is actually Sasaki--Einstein. (For more details see for example
refs.~\cite{hartnoll} and~\cite{BGreview}.)

The Killing condition means that locally we have
\begin{equation}
   K = \dd\psi'+\sigma
\end{equation}
where $\dd\sigma=2mJ$ and that we can write the metric in the form
\begin{equation}
\label{genSE}
   g_X = \hat{g} + K \otimes K
\end{equation}
where $\hat{g}$ is a positive curvature K\"ahler--Einstein
 metric. Note that, by definition, the
metric cone over $g_X$ is Calabi--Yau. All these results generalize
without modification to $(2k+1)$-dimensional Sasaki--Einstein
manifolds $X$.

Finally note that one can group Sasaki--Einstein manifolds by the
nature of the orbits of the Killing vector $K$. If the orbits close,
then we have a $U(1)$ action. Since $K$ is nowhere vanishing, it
follows that the isotropy groups of this action are all finite. Thus
the space of leaves of the foliation will be a positive curvature
K\"ahler--Einstein orbifold of complex dimension $k$. Such
Sasaki--Einstein manifolds are called quasi-regular. If the $U(1)$
action is free, the space of leaves is actually a
K\"ahler--Einstein manifold and the Sasaki--Einstein manifold is
then said to be regular. Moreover, the converse is true: there is
a Sasaki--Einstein structure on the total space of a certain
$U(1)$ bundle over any given K\"ahler--Einstein manifold of
positive curvature~\cite{Kob}. A similar result is true in the
quasi-regular case~\cite{sas-rev}. If the orbits of $K$ do not
close, the Sasaki--Einstein manifold is said to be irregular.

Although there are many results in the literature on
Sasaki--Einstein manifolds explicit metrics are rather rare.
Homogeneous regular Sasaki--Einstein manifolds are classified:
they are all $U(1)$ bundles over generalized flag
manifolds~\cite{tri}. This result follows from the classification
of homogeneous K\"ahler--Einstein manifolds. Inhomogeneous
K\"ahler--Einstein manifolds are known to exist and so one may
then construct the associated regular Sasaki--Einstein manifolds.
However, until recently, there have been no known explicit
inhomogeneous simply-connected\footnote{One can obtain
quasi-regular geometries rather trivially by taking a quotient of
a regular Sasaki--Einstein manifold by an appropriate finite
freely-acting group. Our definition of Sasaki-Einstein will always
mean simply-connected.}
 manifolds in the quasi-regular class. Moreover, no irregular examples
were known at all.

The family of solutions we construct in the following thus gives
not only the first explicit examples of inhomogeneous
quasi-regular Sasaki--Einstein manifolds, but also the first
examples of irregular geometries.

%%%%%%%%%%%%%%%%%%%%%%%%%%%%%%%%%%%%%%%%%%%%%%%%%%%%%%%%%%%%%%%%%%%%%%%%%%

\subsection{The local metric}
\label{sec:SEg}

Let $B$ be a (complete) $2n$-dimensional positive curvature
K\"ahler--Einstein manifold, with metric $g_B$ and K\"ahler form
$J_B$ such that $\Ricci{g_B}=\lambda g_B$ with $\lambda>0$. It is
thus necessarily compact~\cite{Myers} and
simply-connected~\cite{Kob}. We construct the local
Sasaki--Einstein metric~\eqref{genSE} in two steps. First,
following refs.~\cite{BB} and \cite{PP}, consider the local
$2n+2$-dimensional metric
\begin{equation}
\label{metric}
   \hat{g} = \rr^2 g_B
      + U^{-1}\dd\rr\otimes\dd\rr
      + \rr^2 U(\dd\tau-A)\otimes(\dd\tau-A)
\end{equation}
where
\begin{equation}
\label{Udef}
   U(\rr) = \frac{\lambda}{2n+2}
      - \frac{\Lambda}{2n+4}\rr^2
      + \frac{\Lambda}{2(n+1)(n+2)} \left(\frac{\lambda}
         {\Lambda}\right)^{n+2}\frac{\kappa}{\rr^{2n+2}}
\end{equation}
$\kappa$ is a constant, $\Lambda>0$ and
\begin{equation}
   \dd A = 2 J_B ,
\end{equation}
or, in other words, we can take $A=2P_B/\lambda$ where $P_B$ is the
canonical connection defined by $J_B$. By construction $\hat{g}$ is a
positive curvature K\"ahler--Einstein metric with
\begin{equation}
   \hat J = \rr^2 J_B + \rr(\dd\tau - A)\wedge\dd\rr
\end{equation}
and $\Ricci{\hat{g}}=\Lambda\hat{g}$. (Clearly $\hat J$ is closed.
If we let $\hat\Omega$ be the corresponding $(n+1,0)$ form, then
we calculate $\dd \hat\Omega=i\hat P\wedge\hat\Omega$, leading to a
Ricci-form given by $\hat{\cal R}\equiv \dd \hat P=\Lambda \hat J$.)

In ref.~\cite{PP} it was shown that the local
expression~\eqref{metric} describes a complete metric on a
manifold if and only if $\kappa=0$, $B$ is $\CP^n$ and the total
space is $\CP^{n+1}$ the latter each with the canonical metric.
Here we consider adding another dimension to the metric above --
specifically, the local Sasaki--Einstein direction. We define the
$(2n+3)$-dimensional local metric, as in~\eqref{genSE}
\begin{equation}
\label{ES}
   g_X = \hat{g} + (\dd\psi' + \sigma)\otimes(\dd\psi' + \sigma)
\end{equation}
where $\dd\sigma = 2\hat{J}$. As is well-known (see for example
ref.~\cite{hartnoll} for a recent review), such a metric is
locally Sasaki--Einstein. The curvature is $2n+2$, provided
$\Lambda=2(n+2)$. An appropriate choice for the connection
one-form $\sigma$ is
\begin{equation}
   \sigma = \frac{\lambda}{\Lambda} A +
      \left(\frac{\lambda}{\Lambda}-\rr^2\right) (\dd\tau - A) .
\end{equation}
By a rescaling we can, and often will, set $\lambda=2$.

%%%%%%%%%%%%%%%%%%%%%%%%%%%%%%%%%%%%%%%%%%%%%%%%%%%%%%%%%%%%%%%%%%%%%%%%%%

\subsection{Global analysis}
\label{sec:SE-global}

We next show that the metrics~\eqref{ES} give an infinite family
of complete, compact Sasaki--Einstein metrics on a
$2n+3$-dimensional space $X$. Topologically $X$ will be given by
$S^1$ bundles over $\bbP(\mathcal{O}\oplus\mathcal{L}_B)$ where
$\mathcal{O}$ is the trivial bundle and $\mathcal{L}_B$ the
canonical bundle on $B$. However, it should be noted that the
complex structure of the Calabi-Yau cone is \emph{not} compatible
with that on $\bbP(\mathcal{O}\oplus\mathcal{L}_B)$ -- we use the
latter notation only as a convenient way to represent the
topology.

The first step is to make a very useful change of coordinates
which casts the local metric~\eqref{ES} into a different
$(2n+2)+1$ decomposition. Define the new coordinates
\begin{equation}
   \alpha = - \tau - \frac{\Lambda}{\lambda}\psi^{\prime}~
\end{equation}
and $(\Lambda/\lambda)\psi^{\prime} = \psi$. We then have
\begin{multline}
\label{new}
   g_X = \rr^2 g_B + U^{-1}\dd\rr\otimes\dd\rr
      + q (\dd\psi + A)\otimes(\dd\psi + A) \\
      + w (\dd\alpha + C)\otimes(\dd\alpha + C)
\end{multline}
where
\begin{equation}
\label{deff}
\begin{aligned}
   q(\rr) &=
      \frac{\lambda^2}{\Lambda^2}\frac{\rr^2U(\rr)}{w(\rr)} \\
   w(\rr) &= \rr^2 U(\rr)+(\rr^2-\lambda/\Lambda)^2 \\
   C &=
      f(r)
        (\dd\psi +A).
\end{aligned}
\end{equation}
and
\begin{equation}
f(r)\equiv\frac{\rr^2(U(\rr)+\rr^2-\lambda/\Lambda)}{w(\rr)}~.
\end{equation}

The metric is Riemannian only if $U\geq 0$ and hence $w\geq0$ and
$q\geq0$. This implies that we choose the range of $\rr$ to be
\begin{equation}
   \rr_1 \leq \rr \leq \rr_2
\end{equation}
where $\rr_i$ are two appropriate roots of the equation
$U(\rr)=0$. As we want to exclude $\rr=0$, since the metric is
generically singular there, we thus take $\rr_i$ to be both positive
(without loss of generality). Considering the roots of $U(\rr)$ we
see that we need only consider the range
\begin{equation}
   -1 < \kappa \leq 0
\end{equation}
so that
\begin{equation}
   0 \leq \rr_1 < \sqrt{\tfrac{\lambda}{\Lambda}}
      < \rr_2 \leq \sqrt{\tfrac{\lambda (n+2)}{\Lambda(n+1)}} .
\end{equation}
The limiting value $\kappa=0$ in~\eqref{metric} gives a smooth
compact K\"ahler--Einstein manifold if and only if $B=\CP^n$, in
which case $X$ is $S^n$ (or a discrete quotient thereof). As a
consequence, we can focus on the case where the range of $\kappa$
is $-1<\kappa<0$.

Topologically we want $X$ to be a $S^1$ fibration over
$Y=\bbP(\mathcal{O}\oplus\mathcal{L}_B)$
\begin{equation}
\label{SEfibration}
   \begin{CD}
      S_\alpha^1 @>>> X \\
      && @VVV \\
      && Y
   \end{CD}
   \qquad \qquad
   \begin{CD}
      \CP^1_{\rr,\psi} @>>> Y \\
      && @VVV \\
      && B
   \end{CD}
\end{equation}
As indicated, we construct the $S^1$ fibre from the $\alpha$
coordinate and the $\CP^1$ fibre of the base bundle $Y$ from the
$(\rr,\psi)$ coordinates, where $\psi$ is the azimuthal angle and
$\rr=\rr_i$ are the north and south poles.

To make this identification we first need to check that the metric
is smooth at the poles $\rr=\rr_i$. It is easy to show that this
is true provided $\psi$ has period $2\pi$ (with $\lambda=2$ and
$\Lambda=2(n+2)$).

Next consider the $X\to Y$ circle fibration. Suppose $\alpha$ has
period $2\pi\ell$. The question is then, can we choose $\ell$ and
the parameter $\kappa$ such that $g_X$ is a regular globally
defined metric on $X$. The only point to check is that the term
$C$ in the expression for the metric~\eqref{new} is in fact a
connection on a $U(1)$ bundle. The necessary and sufficient
condition is that the periods of the corresponding curvature are
integral, that is
\begin{equation}
\label{Chern}
   \frac{1}{2\pi}F = \frac{1}{2\pi\ell}\dd C
      \in H^2_\text{de Rham}(Y,\bbZ)~.
\end{equation}
(Note that since $B$ is simply connected, so is $Y$ and hence also
$H^2(Y,\bbZ)$ is torsion free. Thus, the periods of $F$ in fact
completely determine the $U(1)$ bundle.)

To check the condition~\eqref{Chern} we need a basis for the
torsion-free part of $H_2(Y,\bbZ)$. First note that since $Y$ is a
projectivised bundle over $B$, we can use the results of sec.~20
of ref.~\cite{bott} to write down the cohomology ring of $Y$ in terms
of $B$. In particular, we have $H^2(Y,\bbZ)\cong\bbZ\oplus
H^2(B,\bbZ)$, where the first factor is generated by the
cohomology class of the $S^2$ fibre. Let $\{\Sigma_i\}$ be a set
of two-cycles in $B$ such that the homology classes $[\Sigma_i]$
generate the torsion-free part of $H_2(B,\bbZ)$. Next define a
submanifold $\Sigma\cong S^2$ of $Y$ corresponding to the fibre of
$Y$ at some fixed point on the base $B$. Finally we also have the
global section $\sigma^N:B\rightarrow Y$ corresponding to the
``north pole'' ($\rr=\rr_1$) of the $S^2$ fibres. Together we can
then construct the set $\{\Sigma,\sigma^N\Sigma_i\}$ which forms a
representative basis generating the free part of $H_2(Y,\bbZ)$.

Calculating the periods of $F$ we find
\begin{equation}
\label{periods}
\begin{aligned}
   \int_{\Sigma} \frac{F}{2\pi}
      &= \frac{f(\rr_1)-f(\rr_2)}{\ell} , \\
   \int_{\sigma^N\Sigma_i} \frac{F}{2\pi}
      &= \frac{f(\rr_2)c_{(i)}}{\ell} ,
\end{aligned}
\end{equation}
where
\begin{equation}
   c_{(i)} = \int_{\Sigma_i}\frac{\dd A}{2\pi}
      =\left<c_1(\mathcal{L}_B),[\Sigma_i]\right> \in \bbZ
\end{equation}
are the periods of the canonical bundle $\mathcal{L}_B$. Thus we have
integral periods if and only if $f(\rr_1)/f(\rr_2)=p/q\in\mathbb{Q}$
is rational with $p,q\in\mathbb{Z}$ and
$\ell=f(\rr_2)/q=f(\rr_1)/p$. The periods of $\frac{1}{2\pi}F$ are
then $\{p-q, qc_{(i)}\}$. Rescaling $\ell$ by
$h=\mathrm{hcf}\{p-q,qc_{(i)}\}$ gives a special class of solutions
where the integral periods $\{h^{-1}(p-q),h^{-1}qc_{(i)}\}$ have no
common factor. This is the class we will concentrate on from now on
since in that case $X$ is simply-connected (see
refs.~\cite{se,gse}).

Notice that, using the expression~\eqref{deff}, we have
\begin{equation}
   R(\kappa)\equiv\frac{f(\rr_1)}{f(\rr_2)}
      = \frac{\rr^2_1(\rr_2^2 - \lambda/\Lambda)}
         {\rr^2_2(\rr^2_1-\lambda/\Lambda)} .
\end{equation}
This is a continuous function of $\kappa$ in the interval
$(-1,0]$. Moreover, it is easy to see that $R(0)=0$ and
$R(-1)=-1$. Hence there are clearly a countably infinite number of
values of $\kappa$ for which $R(\kappa)$ is rational and equal to
$p/q$, with $|p/q|< 1$, and these all give complete Riemannian
metrics $g_X$.

Locally, by construction, $g_X$ was a Sasaki--Einstein metric. In the
$(\alpha,\psi)$ coordinates the unit-norm Killing vector $K$ is given
by
\begin{equation}
    K = \frac{\Lambda}{\lambda}\left(
       \frac{\del}{\del\psi}-\frac{\del}{\del\alpha}\right) .
\end{equation}
Given the topology of $X$ it is easy to see that this is globally
defined. Hence, so is the corresponding one-form and also
$J=\frac{1}{2}\dd K$. Thus the Sasaki--Einstein structure is globally
defined.

Recall that our original problem was to find $X$ which admitted
Killing spinors. For this we need $X$ to be spin. However, by
construction this is the case irrespective of whether $B$ is spin or
not (see ref.~\cite{gse}). Since $X$ is also simply-connected we
can then invoke theorem~3 of~\cite{fried} to see that this implies we
have global Killing spinors.

Finally we notice that the orbits of the Killing vector $K$ close if
and only if $f(\rr_2)\in\mathbb{Q}$, in which case the Sasaki--Einstein
manifold is quasi-regular. For generic $p$ and $q$ the space will be
irregular. Determining when $f(\rr_2)\in\mathbb{Q}$ seems to be a
non-trivial number-theoretic problem. (Though for the case $n=1$ studied
in ref.~\cite{se} one has to solve a quadratic diophantine, which can
be done using standard methods.)  Thus for the countably infinite
number of values of $\kappa$ found here, the K\"ahler--Einstein
``base'' is at best an orbifold, and in the irregular case there is in
fact no well-defined base at all.

%%%%%%%%%%%%%%%%%%%%%%%%%%%%%%%%%%%%%%%%%%%%%%%%%%%%%%%%%%%%%%%%%%%%%%%%%%

\subsection{Five-dimensional solutions}
\label{sec:SE5d}

Of particular interest in string theory are five-dimensional
Sasaki--Einstein solutions. For our class these are the backgrounds
which are dual to the $S^2\times T^2$ solutions in $d=11$ supergravity
discussed in the previous section.

To make the correspondence explicit we take $n=1$, $\lambda=2$ and
$\Lambda=6$ and introduce the new coordinate
$\rr^2=(\lambda/\Lambda)(1-cy)$. The metric~\eqref{new} then takes
the form
\begin{multline}
\label{new2}
   g_X = \frac{1}{6}(1-cy) \tilde{g} + w^{-1}q^{-1}\dd y\otimes\dd y
      + \frac{q}{9} (\dd\psi + \tilde{P})\otimes(\dd\psi + \tilde{P}) \\
      + w (\dd\alpha + C)\otimes(\dd\alpha + C)
\end{multline}
where
\begin{equation}
\begin{aligned}
   q(y) &= \frac{a-3y^2+2cy^3}{a-y^2} , \\
   w(y) &= \frac{2(a-y^2)}{1-cy} , \\
   C &= \frac{ac-2y+cy^2}{6(a-y^2)}(\dd\psi + \tilde{P}).
\end{aligned}
\end{equation}
with $\tilde{g}$ the canonical metric on $S^2$ with
$\Ric_{\tilde{g}}=\tilde{g}$ and $\tilde{P}$ the corresponding
canonical connection. This is the form of the metric one obtains by
making an explicit duality from the $d=11$ solutions given in the
previous section.

Topologically these spaces are all $S^2\times S^3$~\cite{se}. One
finds both quasi-regular and irregular solutions depending on
whether or not $f(\rr_2)\in\bbQ$. In the dual field theory this
corresponds to rational or irrational R-charges and also central
charge. Interestingly the irrational charges are at most quadratic
algebraic, that is can be written in terms of square-roots of
rational numbers. This matches a field theory argument due to
Intriligator and Wecht~\cite{Intriligator:2003jj}.

%%%%%%%%%%%%%%%%%%%%%%%%%%%%%%%%%%%%%%%%%%%%%%%%%%%%%%%%%%%%%%%%%%%%%%%%%%

\subsection{A simple generalisation}
\label{sec:SEgen}

We end by noting that there is a simple generalisation of the
above construction to the case when the base manifold $B$ is a
product of K\"ahler--Einstein manifolds. 

More specifically, let $g_i$ with $i=1,\dots,p$ be a  set of
$2n_i$-dimensional positive curvature K\"ahler--Einstein metrics with
K\"ahler forms $J_i$ and $\Ricci{g_i}=\lambda_ig_i$. There is then a
straightforward generalisation of the construction of refs.~\cite{BB}
and~\cite{PP} allowing one to build a  $(2n+2)$-dimensional
K\"ahler--Einstein metric $\hat{g}$ where $n=\sum_in_i$. In analogy
with~\eqref{metric} we write 
\begin{equation}
\label{gen-metric}
   \hat{g} =  \sum_i \left(r^2+\frac{\lambda_i}{\Lambda}\right) g_i 
      + V(r)^{-1}\dd r\otimes\dd r
      + r^2 V(r) (\dd\tau-A)\otimes(\dd\tau-A)
\end{equation}
and define the corresponding fundamental form
\begin{equation}
   \hat{J} = \sum_i \left(r^2+\frac{\lambda_i}{\Lambda}\right) J_i 
      + r(\dd\tau - A)\wedge\dd r .
\end{equation}
The metric is K\"ahler--Einstein with $\Ricci{\hat{g}}=\Lambda\hat{g}$
provided
\begin{equation}
\begin{aligned}
   \dd A &= 2 \sum_i J_i , \\
   r^2 V(r) &= - \frac{1}{2}\Lambda\,
      f(r^2;\lambda_i/\Lambda,\mu) ,
\end{aligned}
\end{equation}
where
\begin{equation}
   f(x;d_i,\mu) =
      \frac{\mu+\int_0^x \dd x' x' \prod_i (x'+d_i)^{n_i}}
         {\prod_i (x+d_i)^{n_i}} .
\end{equation}
and $\mu$ is an integration constant. Note that we have chosen a
slightly different convention for the coordinate $r$ as compared
to the case of $p=1$ given in~\eqref{Udef}. They differ by
$r^2=\rho^2-\lambda/\Lambda$.

Using the construction of sec.~\eqref{sec:SEg} above we can then
obtain $(2n+3)$-dimensional metrics which are locally
Sasaki--Einstein.  For certain values of the constants $\lambda_i$
and $\mu$ one expects that these will give metrics on complete compact
Sasaki--Einstein manifolds: the analysis will be a direct
generalisation of that in ref.~\cite{gse} but we leave the details for
future work. 

The case of most interest for M-theory is when the dimension of
the resulting Sasaki-Einstein manifold is seven, as they can be
used to obtain new $\AdS_4\times X$ solutions of $d=11$
supergravity. It was shown in ref.~\cite{gse} that the construction
using~\eqref{metric} in case where the base is a direct product of two
equal radius two-spheres generalises the well-known 
homogeneous Sasaki-Einstein metric $Q^{1,1,1}$ (see for example
ref.~\cite{duffrev} for a review). A further generalisation to the
case where the spheres have different radii can be obtained from the
generalised contruction. Setting $p=2$, $n_1=n_2=1$ (so that $g_i$ are
round metrics on two-spheres) in~\eqref{gen-metric} we obtain
\begin{equation}
   \hat{g} = \frac{1}{\Lambda}\left[
      (1+c_1 x)\lambda_1 g_1 + (1+c_2x)\lambda_2 g_2\right]
   + \frac{\dd x\otimes\dd x}{F(x)}
   + \frac{F(x)}{\Lambda^2}(\dd \beta-c_iA_i)\otimes(\dd \beta-c_iA_i) ,
\end{equation} 
where $c_i=\Lambda/\lambda_i$, $\dd A_i=\lambda_i J_i$ and 
\begin{equation}
   F(x) = -\frac{\Lambda}{8}
      \frac{16c_1c_2\mu+8x^2+\frac{16}{3}(c_1+c_2)x^3+4c_1c_2x^4}
         {(1+c_1x)(1+c_2x)} .
\end{equation}

\section*{Acknowledgements}
JFS is supported by NSF grants DMS-0244464, DMS-0074239 and
DMS-9803347. DJW is support by a Royal Society University Research
Fellowship.

%%%%%%%%%%%%%%%%%%%%%%%%%%%%%%%%%%%%%%%%%%%%%%%%%%%%%%%%%%%%%%%%%%%%%%%%%%

%%%%%%%%%%%%%%%%%%%%%%%%%%%%%%%%%%%%%%%%%%%%%%%%%%%%%%%%%%%%%%%%%%%%%%%%%%

\end{document}